\documentclass[12pt,preprint]{aastex}

\usepackage{graphicx}
\usepackage{fancyhdr}
\usepackage{epsfig}
\usepackage{psfrag}
\usepackage{amssymb}
\usepackage{longtable}
\usepackage{epsf}
\usepackage{apjfonts}
\usepackage{mathptmx}
\usepackage{natbib}
\usepackage{lscape}
\usepackage{verbatim}
\usepackage{hyperref} 

\fancypagestyle{plain}{%
  \fancyhf{}
  \fancyfoot[C]{\iffloatpage{}{\thepage}}
  }
\newcommand{\srcno}{61\/}
\newcommand{\fvno}{39\/}
\newcommand{\chandra}{{\it Chandra\/}}
\newcommand{\asca}{{\it ASCA\/}}
\newcommand{\rosat}{{\it ROSAT\/}}

\newcommand{\xmm}{{\it XMM-Newton\/}}
\newcommand{\einstein}{{\it Einstein\/}}

\newcommand{\lum}{\thinspace\hbox{$\hbox{erg}\thinspace\hbox{s}^{-1}$}}
\newcommand{\flux}{\thinspace\hbox{$\hbox{erg}\thinspace\hbox{cm}^{-2}\thinspace\hbox{s}^{-1}$}}
\begin{document}

\def\spose#1{\hbox to 0pt{#1\hss}}
\def\laeq{\mathrel{\spose{\lower 3pt\hbox{$\mathchar"218$}}
     \raise 2.0pt\hbox{$\mathchar"13C$}}}
\def\gaeq{\mathrel{\spose{\lower 3pt\hbox{$\mathchar"218$}}
     \raise 2.0pt\hbox{$\mathchar"13E$}}}

\title{LONG-TERM X-RAY VARIABILITY STUDY OF IC342 FROM XMM-NEWTON OBSERVATIONS}

\author{Daisy S.Y.~Mak\footnote{current affilation: Department of Physics and Astronomy, University of Southern California, Los Angeles, CA, 90089--0484, US}}
\affil{Department of Physics, University of Hong Kong, Pokfulam Road, Hong Kong, PR China}
\author{Chun S.J.~Pun}
\affil{Department of Physics, University of Hong Kong, Pokfulam Road, Hong Kong, PR China}
\author{Albert K.H.~Kong\footnote{Golden Jade Fellow of Kenda Foundation,
Taiwan}}
\affil{Institute of Astronomy and Department of Physics, National Tsing Hua University, Hsinchu, Taiwan 30013, R.O.C.}

\begin{abstract}
We presented the results of an analysis of four \xmm\ observations of the starburst galaxy IC342 taken over a four-year span from 2001 to 2005, with an emphasis on investigating the long-term flux and spectral variability of the X-ray point sources. We detected a total of \srcno\ X-ray sources within $35'\times30'$ of the galaxy down to a luminosity of $(1 - 2)\times10^{37}\lum$ depending on the local background. We found that \fvno\ of the \srcno\ detected sources showed long-term variability, in which 26 of them were classified as X-ray transients. We also found 19 sources exhibiting variations in hardness ratios or undergoing spectral transitions among observations, and were identified as spectral variables. In particular, 8 of the identified X-ray transients showed spectral variability in addition to flux variability. The diverse patterns of variability observed is indicative of a population of X-ray binaries. We used X-ray colors, flux and spectral variability, and in some cases the optical or radio counterparts to classify the detected X-ray sources into several stellar populations. We identified a total of 11 foreground stars, 1 supersoft sources (SSS), 3 quasisoft sources (QSS), and 2 supernova remnants (SNR). The identified SSS/QSS are located near or on the spiral arms, associate with young stellar populations; the 2 SNR are very close to the starburst nucleus where current star formation activities are dominated. We also discovered a spectral change in the nuclear source of IC342 for the first time by a series of X-ray spectrum analysis.

\end{abstract}


\section{INTRODUCTION}
IC342 is a nearby (1.8Mpc; see~\citealt{Buta1999} for a review) late-type Sc/Scd galaxy in the Maffei Group which is one of the closest groups to our Galaxy. Its spiral arms are well developed and are almost face-on ($i = 25^{\circ}\pm3^{\circ}$;~\citealt{Newton1980}). Its active star formation activities in the nuclear region has made it a popular target for infrared and submillimeter observations (e.g.~\citealt*{Boker1999},~\citealt{Meier2005}). These observations revealed that the physical properties of molecular clouds, the infrared luminosity, and the presence of a nuclear stellar cluster in IC342 are similar to the Milky Way in many ways~\citep{Schinnerer2003,Schulz2001}. 
Its proximity and its orientation provide a unique possibility to study the X-ray sources in IC342 from a very favorable viewing angle. One major drawback of X-ray study is that this galaxy is located near the galactic plane ($b=10.6^{\circ}$) and has a large absorption ($N_{\rm H}=3\times10^{21}\rm~cm^{-2}$) towards the Galactic center~\citep{Stark1992}. This limits us to constrain local absorption and determine X-ray emission below 1 keV. On the other hand, IC342 has been studied in the X-ray with many missions since \einstein. With the advent of high angular resolution and high sensitivity instruments such as \chandra\ and \xmm\, it is possible to study the X-ray source population in depth.

The first X-ray observation of IC342 from \einstein\ ~\citep{Fabbiano1987} showed that the X-ray emission was dominated by three ultraluminous X-ray sources (ULXs) which had luminosities above $10^{39}\lum$ (named X-1, X-2, and X-3 based on designations of~\citealt{Okada1998}). Subsequent \rosat\ HRI observation in 1991 revealed seven additional point sources in the disk with a detection limit of $L_{0.1-2.5\rm keV}\approx2\times10^{37}\lum$~\citep{Bregman1993}. Two \asca\ observations taken in 1993 and 2000 on X-1 and X-2 ~\citep{Okada1998,Kubota2002} showed that both X-1 and X-2 exhibited spectral and intensity transitions which resembled the spectral/intensity states of X-ray transients in our Galaxy.~\citet{Kong2003} and~\citet{Bauer2003} presented the result from a 2001 \xmm\ 10 ks observation and both reported detections of about 35 sources. The slope of the X-ray luminosity function was found to be $\sim0.5$ in both studies, suggesting the X-ray population of IC342 was consistent with other starburst galaxies and Galactic HMXBs. A detailed analysis on the \chandra\ HRC-I image observed in 2006 was presented in~\citet{Mak2007} in which the nuclear X-ray source (X-3) was found to be spatially coincident with a nuclear stellar cluster which had been identified in earlier optical and infrared observations (e.g.~\citealt{Boker1999}). Thus they proposed that the source was not an ULX and was instead associated with starburst activities, together with the possibility of an embedded low luminosity AGN in the nuclear stellar cluster.

X-ray observations of our own Galaxy revealed a diverse population of sources, including low-mass X-ray binaries (LMXBs), high-mass X-ray binaries (HMXBs), and a few young supernova remnants (SNRs) that show various kinds of variability~\citep{Fabbiano2006}. Variability on flux and spectral properties are useful for characterization of the emission mechanism of X-ray sources. However, not much work had been done on the long-term variability of the X-ray sources in IC342 though there were multiple epochs over the previous two decades.  Recently IC342 was observed four times with \xmm\ between 2001 and 2005, and five times with \chandra\ between 2002 and 2006. Therefore, we made use of archival \xmm\ observations which spanned a period of four years to study the nature of the X-ray population in IC342. 
 
In this paper, we present the source catalog and source properties in the four \xmm\ observations. The observation and the data reduction procedures of the X-ray data are presented in Section~\ref{s:obs}. In Section~\ref{s:ana}, the analysis of the data, including source detection, photometry, computation of hardness ratios of the X-ray sources are described. In Section~\ref{s:res}, we present the results of the identification and variability properties of the detected X-ray sources. We finally discuss the global X-ray source population in IC342, with detailed descriptions of several bright individual sources in Section~\ref{s:pro}. 

\section{OBSERVATIONS AND DATA REDUCTION}
\label{s:obs}
IC342 was observed with the European Photon Imaging Camera (EPIC) on board \xmm\ four times from 2001 to 2005. The instrument modes were full-frame with medium optical blocking filter for the EPIC-PN and the two EPIC-MOS camera. The observation in 2001 was part of a multi-wavelength campaign to study star formation activities of IC 342 and the aimpoint was at the nuclear region. The three subsequent observations were made to study the ULXs in IC 342 with two long observations in 2004 ($T_{\rm exp}\geq20$ ks) aiming at X-1 and one short observation ($T_{\rm exp}\approx6$ ks) aiming at X-2 in 2005. These observations covered the entire galaxy with a circular field of view of $30'$ diameter. Since the field of view of these observations overlapped and were made at various aimpoints, the total region covered was about $35'\times30'$. Therefore sources at the outer edges of the CCD might not be observable in all four exposures.
A summary of these observations was listed in Table~\ref{t:obs}. Even though the 2001 February observation had already been studied in detail by~\citet{Kong2003} and~\citet{Bauer2003}, for for consistency we included it also in our present analysis and reanalysed it whenever necessary. 

The event files were filtered and reprocessed using the \xmm\ Sciences Analysis System (SAS v7.1.0). Time intervals contaminated by soft-proton flares were identified using the background light curves in the $>10$ keV band. High background level with count rate over 0.2 cps for MOS and 1.0 cps for EPIC-PN were excluded and the good time intervals (GTI) were obtained for each observation which ranged from 5.6 to 23.6 ks. The resulting GTI of the two observations in 2004, with exposure time over $\geq20$ ks for each MOS camera, roughly double those of the 2001 and 2005 data. Therefore the sensitivity of the 2004 observations were expected to be higher and fainter sources could be detected. We selected only good event patterns for imaging: $\leq12$ for MOS and $\leq4$ for PN, and restricted our analysis in the energy range 0.2--12 keV. These filtered event files were then used for data analysis using HEAsoft v6.4 and XSPEC v12.4. 

\section{ANALYSIS}
\label{s:ana}
\subsection{Source Detection and Astrometry}
\label{ss:det}
Source detection using maximum likelihood fitting was done simultaneously on each of the EPIC-PN, MOS1, and MOS2 image in the three energy bands: soft (S; 0.2--1.0 keV), medium (M; 1.0--2.0 keV), and hard (H; 2.0--12.0 keV), using the SAS task EDETECT\_CHAIN. A likelihood value of 10 was used, corresponding to a significance level of $3.6\sigma$. The outputs from EDETECT\_CHAIN were inspected for spurious sources at the edges and boundaries between chips. To generate the final source list, we imposed two additional selection criteria for sources detected by EDETECT\_CHAIN for a source to be confirmed as a true detection if it satisfied either one of the followings: First, the source had to be detected in at least one additional observation within a searching radius of $6''$; second, if the source was detected in only one observation, the S/N of the source had to be greater than 5 in that detection. These criteria were applied to all observations, except for the 2001 February observation in which the sources found by EDETECT\_CHAIN were consistent with the source list in~\citet{Kong2003} and~\citet{Bauer2003} and thus we just adopted the source list in~\citet{Kong2003} for this observation. About one third of the sources detected in the four observations were eliminated by these criteria. 

Using this algorithm, we found 37, 43, 51, and 30 sources in the observation taken in 2001 February, 2004 February, 2004 August, and 2005 February respectively. Combining the individual source lists, we identified a total of \srcno\ sources in our final X-ray catalog. As a cross check, we compared the sources detected by the EDETECT\_CHAIN software in SAS with those detected by the WAVDETECT task of CIAO~\citep{Freeman2002}, running on a merged MOS+PN image with a significance threshold of $1\times10^{-6}$. Most sources that were detected with EDETECT\_CHAIN in the final catalog could also be detected by WAVDETECT, except for sources 1, 2, and 53 which were only detected with EDETECT\_CHAIN in the longest exposure data (2004 Auguest). It is possible that they are spurious sources due to peculiar point spread function features or the presence out-of-time event signatures~\citep{Argiroffi2006}. 
On the other hand, these sources might be transients that could be detected during limited period. However, the fact that the three sources are all observed only in the deepest observations could simply mean that they were below detection limit in the other exposures. Therefore we consider them as real detections.

The absolute astrometry of source positions are limited by two factors. Firstly, the aspect solution of \xmm\ is accurate to $\approx4''$. Second, the systematic shifts among the observations made it possible that source detection processes run in different bands and observations might give slightly offset centroids for the same source. In order to determine source positions, we first created a single-stacked image by aligning the four observations in the coordinate frame of the 2004 February image. Eight bright X-ray sources common to all data sets within $10'$ of the aimpoint were selected for the alignment purpose. The relative shifts between these exposures were determined using the IRAF task CCMAP. The cross registration gave a rms accuracy of $0.8''$. Positions of the \srcno\ discrete sources in our final X-ray catalog were then determined in the aligned single-stacked image with WAVDETECT. The positional errors from WAVDETECT varied as the off-axis angle from the center of the image, with sources at the outer edges having larger errors up to $2''$ and these were tabulated in Table~\ref{t:cat}. We attemped to improve the astrometic accuracy by cross-correlating the output X-ray positions from WAVDETECT with the optical USNO-B1.0~\citep{Monet2003} and near IR 2MASS~\citep{Cutri2003} catalogs. We identified 25 X-ray sources with optical counterparts and 16 having near IR counterparts, using a searching radius of $5''$. We selected 7 X-ray sources which possessed consistent and high S/N optical counterparts in at least three observations, to correct for the astrometry of the X-ray images with the optical positions using the IRAF CCMAP task. The average astrometic rms were $1.05''$ and $1.14''$ in RA and DEC respectively. The positions listed in Table~\ref{t:cat} and throughout this paper are based on this astrometric reference. 

\begin{figure}[htbp]
	\begin{center}  
	\leavevmode 
	\parbox{16cm}{  
	\includegraphics[width=16cm]{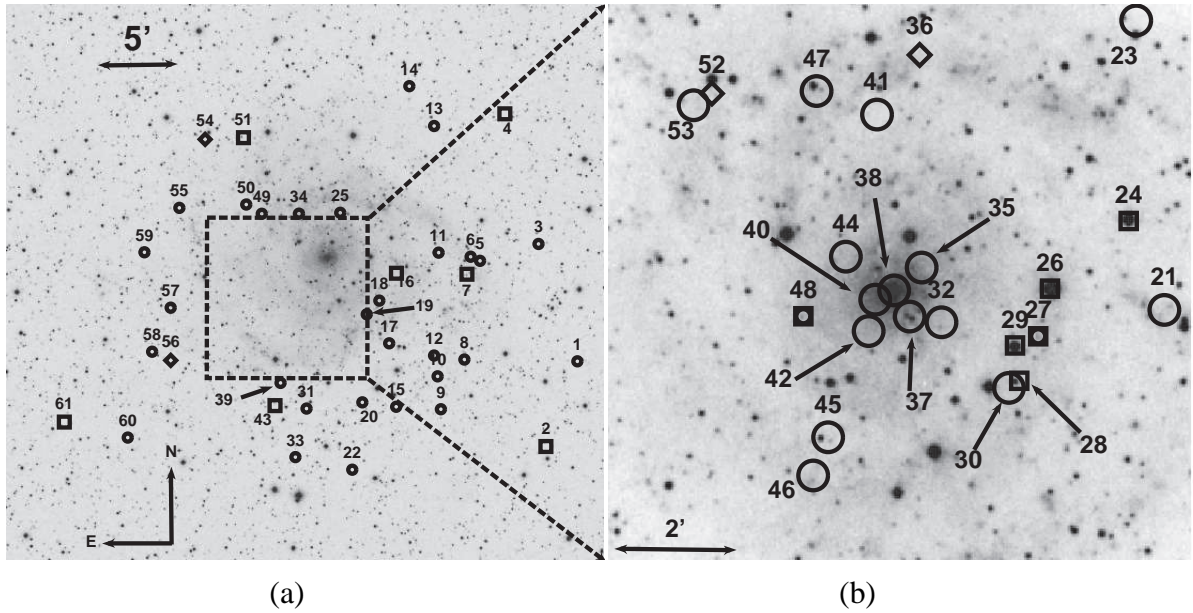} 
	\hspace*{3.5cm}(a)
	\hspace*{7cm}(b) }
\end{center}
	\caption{Digitized Sky Survey (DSS) blue band image of (a) the field of view ($35'\times35'$) of \xmm\ and (b) the central $10'\times10'$ of IC342, with detected \chandra\ X-ray sources overlaid. SSS/QSS (diamonds), foreground stars (squares), SNRs (boxcircles) are identified while the unclassified X-ray point sources (circles) are also shown. The radii of the unclassified X-ray source circles are $15''$. }
	\label{f:ide}
\end{figure}

The spatial distribution of the X-ray sources in IC 342 is not uniform. The majority of the sources detected are located on the spiral arms of the galaxy (Figure~\ref{f:ide}). The 17 sources that are outside the $D_{25}$ disk ($D\sim21.4'$,~\citealt{Vaucouleurs1992}) of IC342 are most likely to be contaminated by foreground or background objects. To estimate the number of background objects within the $D_{25}$ disk, we used the \chandra\ Deep Field data~\citep{Brandt2001} and concluded that there were about 10 background sources based on a detection limit of $f_{\rm X}\sim2.7\times10^{-14}\flux$, corresponding to 23\% of the total number of sources inside the $D_{25}$ disk. This number is consistent with the result by~\citet{Bauer2003} (7--12 based on the detection limit of $f_{\rm X}\sim1.0\times10^{-14}\flux$) but higher than that by~\citet{Kong2003} ($\approx3$ based on the completeness limit of $f_{\rm X}\sim7.7\times10^{-14}\flux$). The difference in the number of background sources estimated from these two groups is probably due to the use of different flux limit values. 

\subsection{Cross-correlation with X-ray catalogs}
\label{sss:tra}
We performed a systematic cross-correlation with existing IC342 catalogs from other X-ray missions including \einstein, \rosat, \chandra, as well as previous \xmm\ publications. Source identification are compared with this new catalog by varying the searching radius according to the spatial resolution of the corresponding X-ray instruments. The results were listed in Table~\ref{t:xso}. The 3 brightest X-ray sources in IC342 ULX-1, ULX-2, ULX-3, first detected by \einstein, were also detected by the current \xmm\ observations. For the \rosat\ sources, all except R2 (designation from~\citealt{Bregman1993}) were detected in our current \xmm\ observations. R2 was the faintest source in the \rosat\ observation with only $14.7\pm5.5$ counts in the 0.1--2.5 keV range, suggesting that it might be too faint to be detected in our observations and was thus excluded in our source detection algorithm. We found 21 of our \xmm\ sources to have counterparts in the \chandra\ HRC-I catalog~\citep{Mak2007}. Four of the \chandra\ sources were not detected in our \xmm\ observations: two (C10, C13; designation from~\citealt{Mak2007}) of them were located close to the nucleus and were not resolved in the \xmm\ data; one (C20, probably a foreground star) of them was outside the field of view; and the remaining one (C19) was either below the detection limit of \xmm\ or might be a variable. We also recovered all the \xmm\ sources previously listed in~\citet{Bauer2003} and~\citet{Kong2003} that was based on the 2001 February observation. The sources lists in both works were very similar, with \citet{Kong2003} listing two more sources than~\citet{Bauer2003}. The X-ray properties, including color-color diagram, spectral fits of bright sources, of the 2001 February observation analyzed in our work were consistent with these two references.

There were 22 \xmm\ sources that were not detected in previous \einstein, \rosat, and even \chandra\ observations, presumably because they were below the detection limit or not resolved by these instruments. It is worth noting that the \xmm\ observations in the present analysis should reach the lowest flux level. For example, the exposure time of the 2004 August \xmm\ observation at 23.6~ks was a factor of $\sim2$ times longer than the \chandra\ HRC-I observation, and 20\% longer than the \rosat\ observation. Nevertheless, it is possible that these 22 sources might be variables and were thus not observed in previous X-ray data. As we will discuss later in section~\ref{ss:var}, 14 of these 22 sources were classified as variables and/or X-ray transients. For example, source 45 was detected in both the 2004 observations and became very luminous, with $L_{\rm X}=4.9\times10^{38}\lum$, but it was dimmer than the detection limit in the 2005 February observation. These comparisons helped to identify the long-term variability of the X-ray sources.

\subsection{Photometry}
\label{ss:cnt}
There are two methods to perform photometry on the X-ray images: one with the SASDAS task EMLDETECT specifically designed for analyzing the \xmm\ data and another one with the CIAO task DMEXTRACT that is compatible with both \xmm\ and \chandra\ data. The count rates and fluxes from the output of EMLDETECT have been corrected for vignetting and instrumental PSF that account for the variation across the CCD before they are background subtracted. However, there is a limitation in using this tool to determine the net source counts since we do not have detailed knowledge of the background subtraction and exposure correction implemented within the tool (M. Ehle, private communication). Therefore some of these source count rates could possibly be contaminated by nearby sources in crowded regions (e.g. the center of IC342). On the other hand, the net source count generated from the task DMEXTRACT do not suffer from this problem. As a result, the net source counts were determined by DMEXTRACT while the count rates and fluxes were determined by EMLDETECT in this work. 

The X-ray fluxes of individual sources were determined in each observation, even when the source was below the detection limit in that particular data set. This was done by running the SASDAS task EMLDETECT with the final position of each source as an input parameter (by setting fitposition=no) and setting the detection threshold zero. The aperture sizes were automatically determined by the maximum likelihood PSF fitting algorithm. Furthermore, to convert the count rates to fluxes, we used the energy conversion factors (ECF) calculated using WEBPIMMS by assuming an absorbed power-law spectrum with a photon index of 2 and an absorption column density of $8\times10^{21}\rm cm^{-2}$ (typical for X-ray Binaries;~\citealt{Kong2003}). The resulting ECF values were listed in Table~\ref{t:ecf} and were used as input parameters for EMLDETECT. This power-law model was used since the X-ray population is dominated by X-ray Binaries and the fact that hardness ratios of most sources in IC342 are consistent with such model (c.f. Section~\ref{ss:hr}). For non-detections, we determined the $3\sigma$ upper limit of the flux and luminosity of the sources assuming a distance of 1.8 Mpc to IC342.

To extract source counts, we produced mosaiced X-ray images from each observations that combined data from the three EPIC cameras in the full band, followed by normalization using the exposure maps created from the SAS tool EEXPMAP. We extracted the source counts via aperture photometry and varied the aperture size for each sources based on their off-axis angle in order to match the 90\% encircled energy function. The extraction radii were smaller near the aimpoint where the PSFs were well defined and bigger at larger off-axis angle where the spatial resolution was poorer. The net counts for each of the three bands listed in Table~\ref{t:cat} are the sum from the 4 observations.

\subsection{Hardness Ratio}
\label{ss:hr}
\begin{figure}[htbp]
\begin{center}
\leavevmode	
	\includegraphics[width=10cm, angle=0]{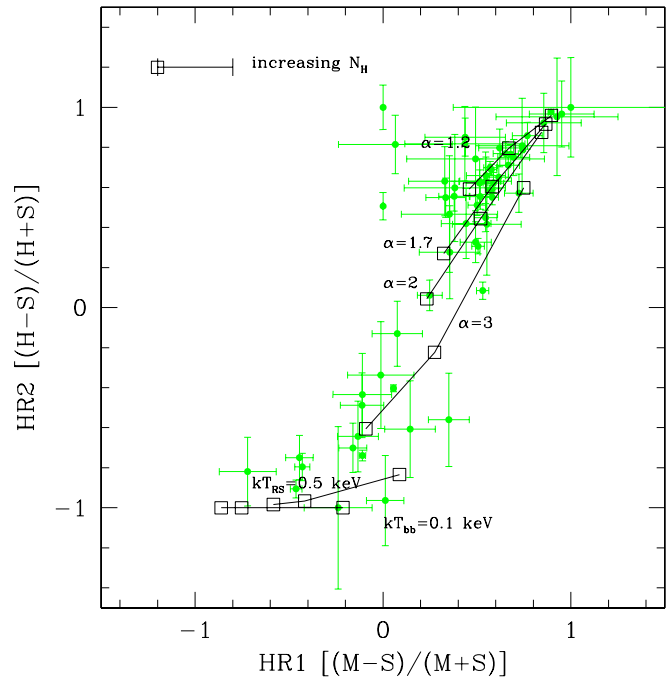}\\
\end{center}
	\caption{Color-color diagram for all sources with more than 20 counts (circles: detected sources; squares: spectral models). Also plotted are the theoretical hardness ratios estimated from different spectral models. \textit{Top to bottom:} Power-law model with $\Gamma$ = 1.2, 1.7, 2, and 3; Raymond-Smith model with $kT_{\rm RS}=0.5$ keV; Blackbody model with $kT_{\rm BB}=0.1$ keV. For each model, $N_{\rm H}$ varies from left as $3\times10^{21}\rm cm^{-2}$, $5\times10^{21}\rm cm^{-2}$, and $10^{22}\rm cm^{-2}$~\citet{Kong2003}.
	\label{f:allhr}}
\end{figure}

Most of the X-ray sources in IC342 are faint with fewer than 200 counts in each observation, which makes it difficult to derive accurate spectral parameters. Hardness ratios can provide crude indications of the X-ray spectra in these cases. We computed the hardness ratios for each detected source, based on the source counts in soft ($S$), medium ($M$), and hard ($H$) bands. Following the convention in~\citet{Kong2003}, we computed two hardness ratios defined as $HR1=(M-S)/(M+S)$ and $HR2=(H-S)/(H+S)$. Table~\ref{t:cat} lists the average values for both HR1 and HR2 over the four observations for each detected sources. Figure~\ref{f:allhr} shows the color-color diagram (CD) for all detected sources. We also checked the CD for each of the four observations individually and found that they were similar. In particular, the CD from the 2001 data is consistent with the results of~\citet{Kong2003}.
We overlaid on the CD six lines showing the tracks followed by representative spectral models with different values of $N_{\rm H}$: power-law models with power-law index $\Gamma$ = 1.2, 1.7, 2, and 3; Raymond-Smith model with $kT_{\rm RS}=0.5$ keV; blackbody model with $kT_{\rm BB}=0.1$ keV. For each model, $N_{\rm H}$ varied from left to right with $3\times10^{21}\rm cm^{-2}$, $5\times10^{21}\rm cm^{-2}$, and $10^{22}\rm cm^{-2}$. The power-law spectra tend to occupy the upper right region of the CD while the soft thermal models tend to occupy the lower left region. Typical X-ray binaries or AGNs with power-law spectrum would most likely be located at the upper right region while supersoft sources (SSS) having blackbody spectra of $kT_{\rm BB}\sim0.1$ keV would occupy at the lower left.

The X-ray sources in IC342 are dominated by sources emitting power-law spectra, with only a few of them compatible with thermal spectra. These sources with soft X-ray emission could possibly be supersoft X-ray sources, supernova remnants, or foreground stars. This trend is consistent in all the four datasets as shown in their individual CDs. To further classify the \xmm\ sources into subcategories such as foreground stars, SNR, and SSS, we employed the hardness ratio selection algorithm by~\citet{Misanovic2006} for the foreground star and SNR classification, and that of~\citet{Stefano2003b} for SSS and Quasisoft sources (QSS) classification. Since the definition of the hardness ratios employed in these publications are slightly different from our chosen energy bands, we computed the corresponding HR values used for each classification scheme. Detailed results will be discussed below in Section~\ref{ss:ide}.

\section{Results}
\label{s:res}
\subsection{Flux Variability}
\label{ss:var}
The four \xmm\ observations analysed in this work spanned a period of four years from 2001 to 2005, allowing us to study the long-term X-ray variability of these sources. Following the definitions by \citet*{Primini1993}, we computed a flux variability parameter $S_{\rm flux}$ defined as
\begin{equation}
S_{\rm flux}=\frac{\left|F_{\rm max}-F_{\rm min}\right|}{\sqrt{\sigma^{2}_{F_{\rm max}}+\sigma^{2}_{F_{\rm min}}}} \ ,
\label{eq:f}
\end{equation}
where $F_{\rm max}$ and $F_{\rm min}$ are the maximum and minimum X-ray full-band fluxes of the four observations with $\sigma^{2}_{\rm F_{max}}$, and $\sigma^{2}_{\rm F_{\rm min}}$ as their corresponding errors. A source is defined to be a flux variable if $S_{\rm flux} > 3$ and is marked with "f" in the last column of Table~\ref{t:cat}. If the minimum flux is a non-detection (i.e. S/N$<5$), we used the $3\sigma$ flux upper limit to calculate the lower limit of the variability factor. Since most of the sources in our catalog have low number of counts with nonzero background, we employed the Bayesian approach to calculate the $3\sigma$ flux upper limit~\citep*{Kraft1991}. 

\begin{figure}[htbp]
\begin{center}
\leavevmode	
	\includegraphics[width=10cm, angle=0]{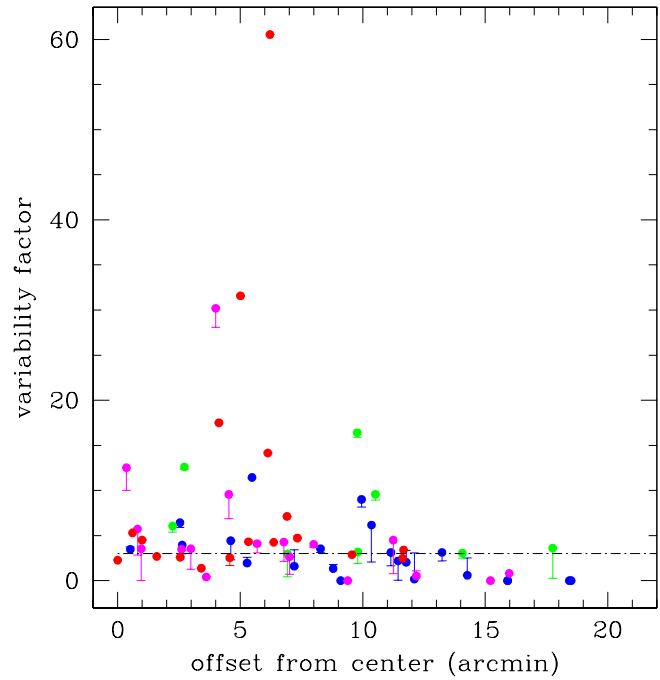}\\
\end{center}
	\caption{The flux variability factor $S_{\rm flux}$ versus the average angle offset from the galactic center of IC342. Blue: sources detected in only one observation; Green: sources detected in two observations; Magenta: sources detected in three observations; Red: sources detected in all 4 observations;
 Sources with $S_{\rm flux}>3$ (above the dash line) are defined to be flux variable. The arrows attached to the data point represents the $3\sigma$ lower limit of the variability factor.
	\label{f:var}}
\end{figure}

Long-term X-ray flux variability were detected for \fvno\ out of \srcno\ sources , representing 64\% of the total. Of these, 29 sources varied in flux by more than a factor of two (i.e. $F_{\rm max}/F_{\rm min}\ge2$). The maximum amount of variabiliy corresponds to $S_{\rm flux}\sim60$. We plotted the relation between the average offset from the galactic center and the variability factor of each source in Figure~\ref{f:var}. It is noted that most of the strong variables are located between $\sim3'-7'$, corresponding roughly to the region of the spiral arms, hinting that these sources are indeed located within the galaxy and are not foreground or background sources.

We have also adopted the criteria of~\citet{Kong2002} to search for X-ray transients which are defined as sources: (1) having flux variability factor $S_{\rm flux} > 3$, and (2) found in at least one observation with $L_{\rm X}\ge10^{37}\lum$ and was not detected (i.e. source counts are below the $3\sigma$ detection threshold) in at least one other observation. The luminosity limit covers typical outburst luminosities of soft X-ray transients and Be/X-ray binaries in our Galaxy. A total of 26 X-ray transients are detected (about 43\% of the total \srcno) and are marked with "t" in Table~\ref{t:cat}. About half of these transients were bright during their high state, $L_{\rm X}\ge3\times10^{37}\lum$, the brightest of which (source 36) has $L_{\rm X}=4.9\times10^{38}\lum$, meaning that the source had brightened by a factor of $> 50$.

\begin{figure}[htbp]
        \parbox{8cm}{
	\includegraphics[bb= 18 144 592 718,width=8cm, angle=0]{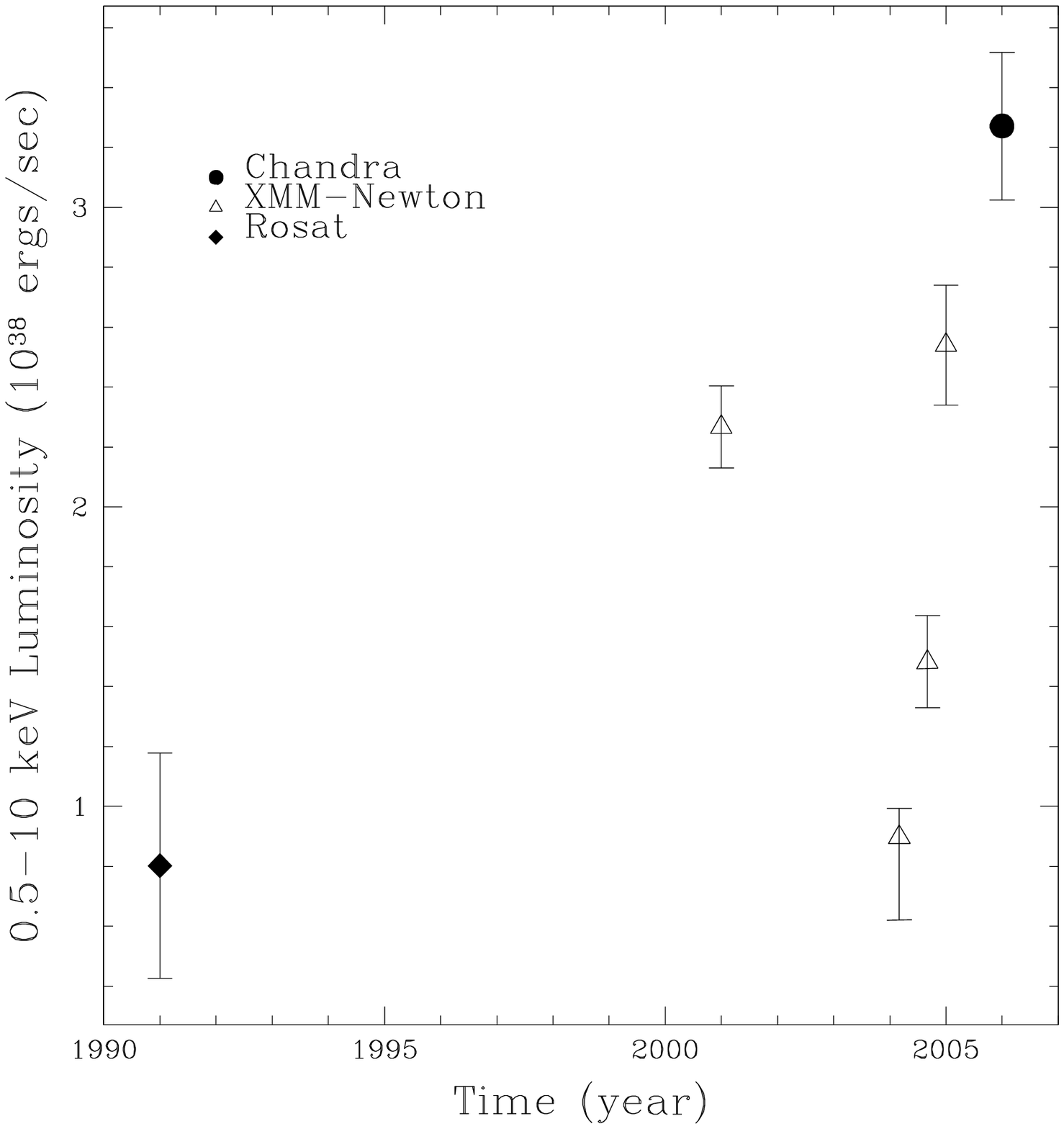} \\
	\hspace*{2.5cm}Source 17}
	\hfill
       \parbox{8cm}{
	\includegraphics[bb= 18 144 592 718,width=8cm, angle=0]{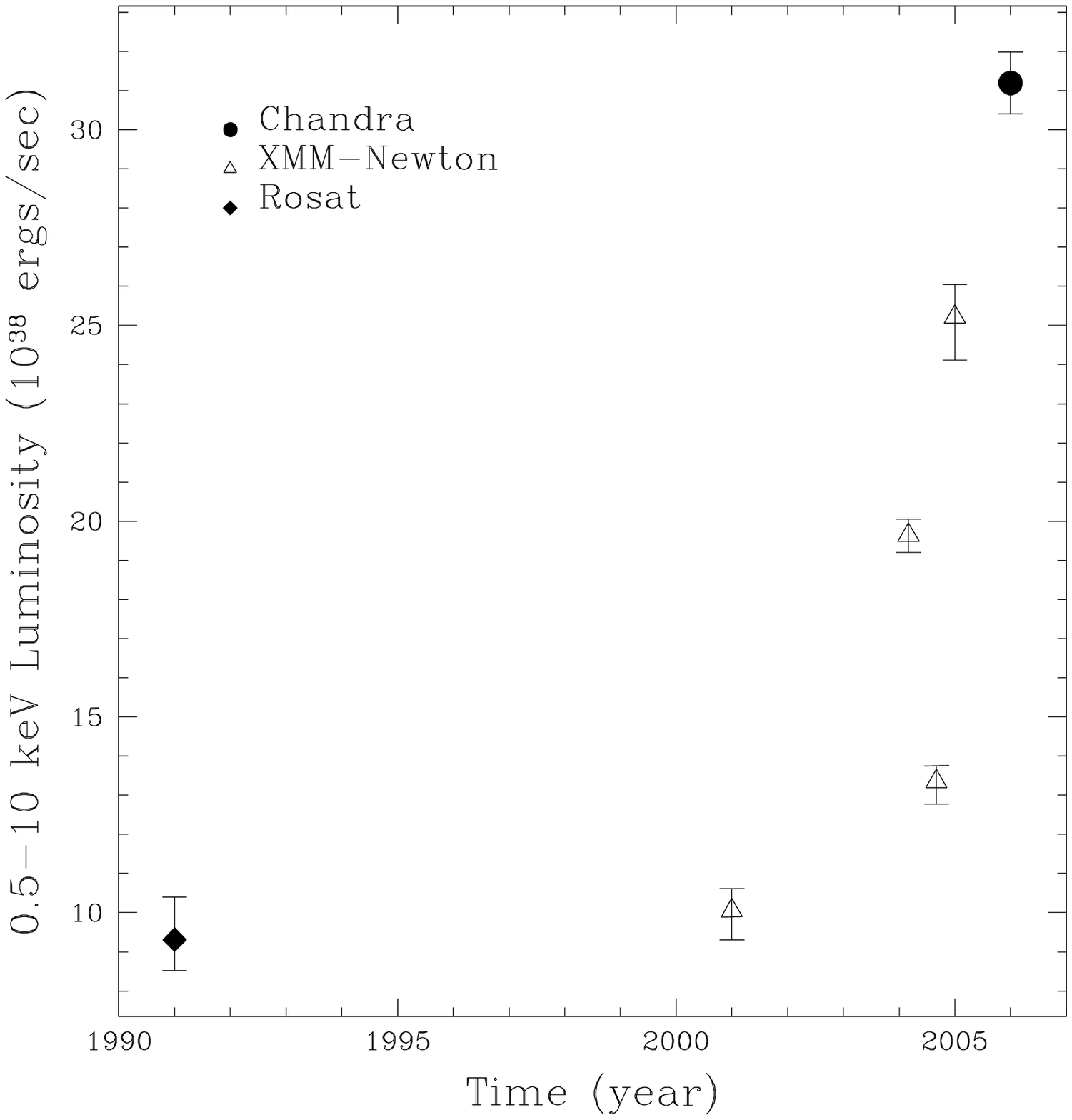}\\
	\hspace*{2.5cm}Source 19 (IC342 X-1) }

        \vspace{0.5cm}
        \parbox{7cm}{
	\includegraphics[bb= 18 144 592 718,width=8cm, angle=0]{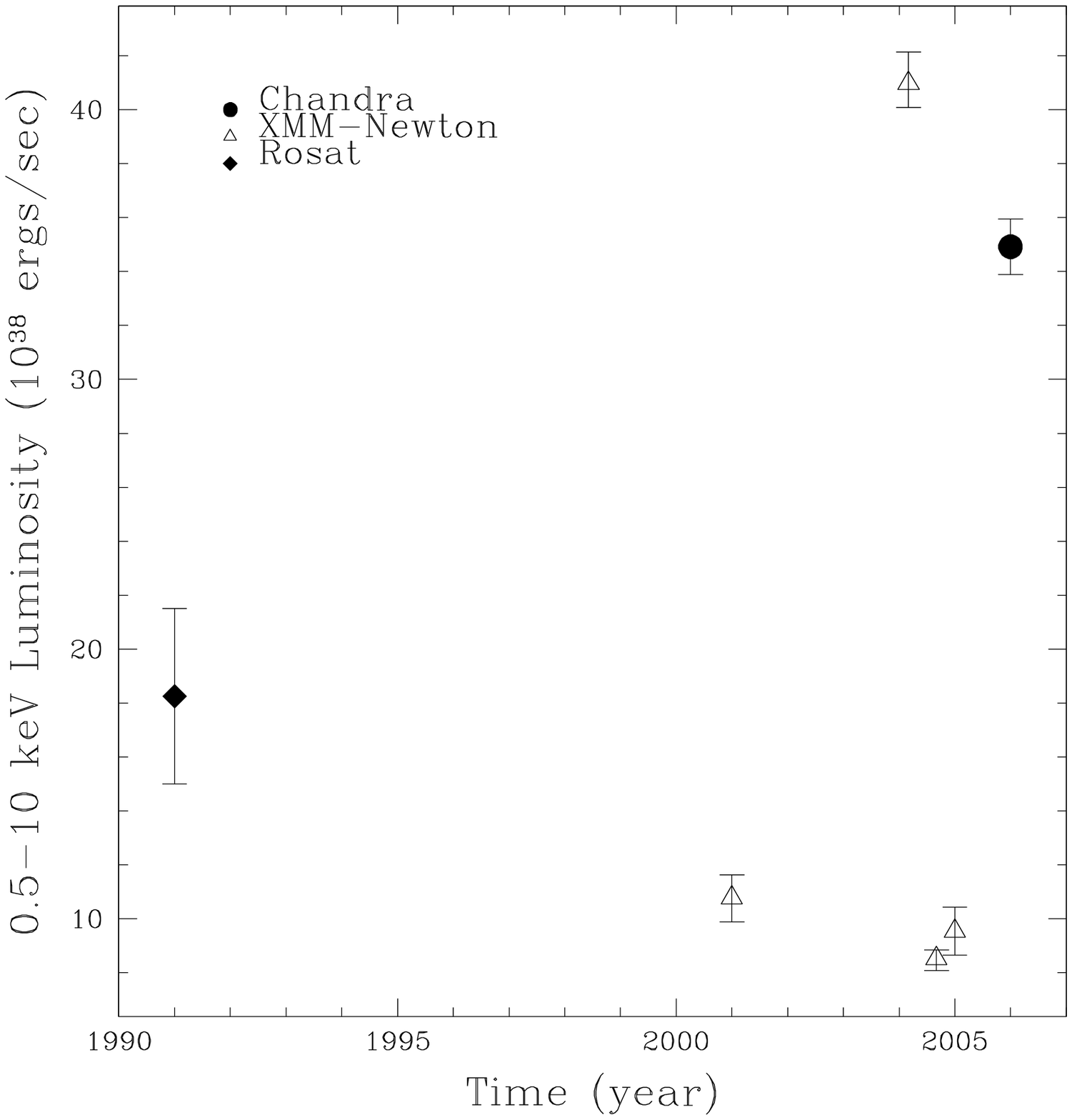}\\
	\hspace*{2.5cm}Source 25 (IC342 X-2 }
	\hfill
        \parbox{8cm}{
	\includegraphics[bb= 18 144 592 718,width=8cm, angle=0]{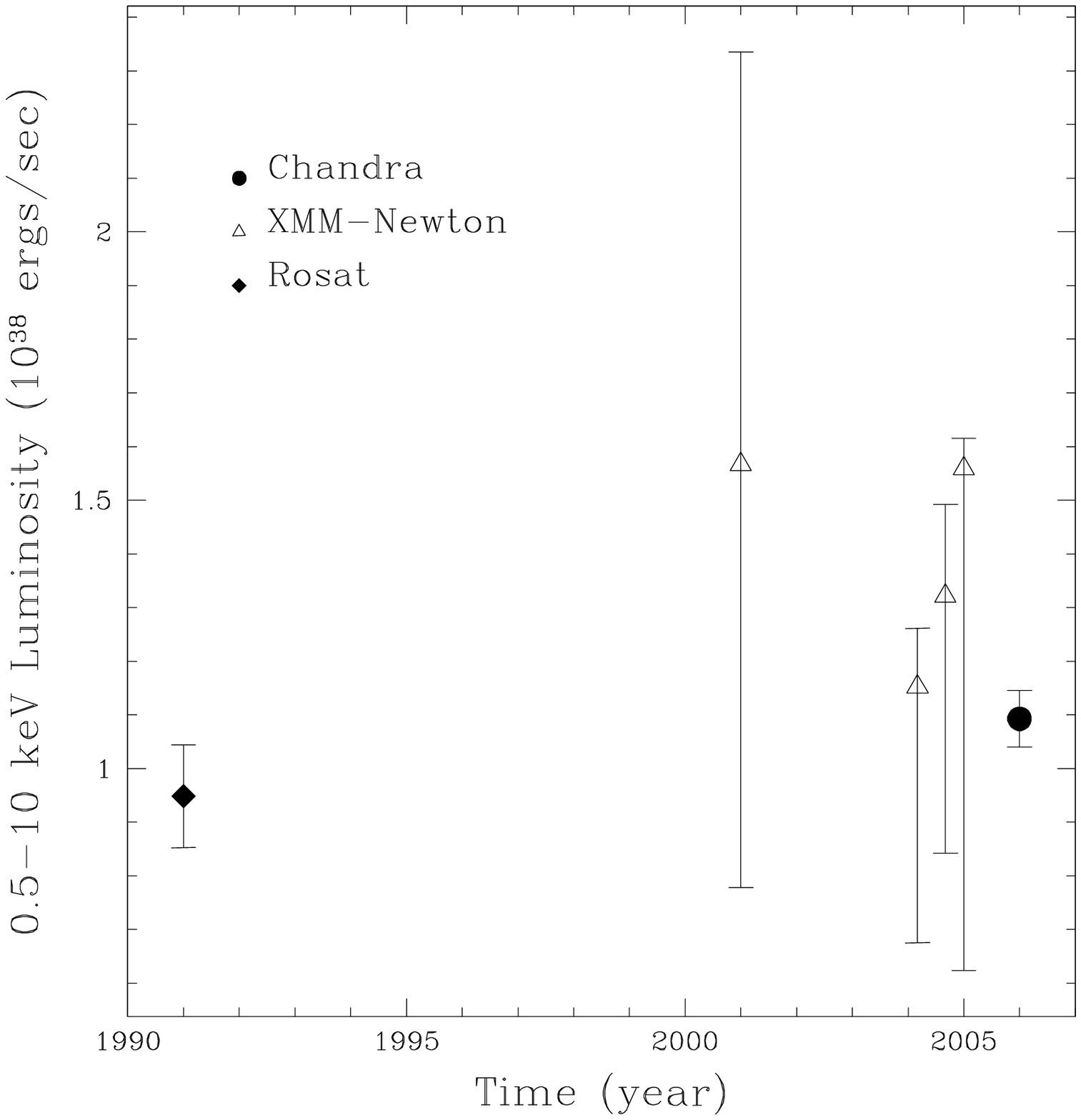}\\
	\hspace*{2.5cm}Source 38 (IC342 nucleus) }
	\caption{Long-term light curve of the X-ray sources in IC342: source 17 (top-right), source 19 (top-left), source 25 (bottom-right), and source 38 (bottom-right). Filled circle represent \chandra\ HRC-I observations, triangles represent \xmm\ EPIC observations, and diamonds represent \rosat\ HRI observations. Errors for \xmm\ data points are $1\sigma$ gaussian limits from XSPEC while for \chandra\ and \rosat\ data points are derived poisson errors of the source counts. It was noted that the \xmm\ and \rosat\ fluxes of source 38 could suffer from background contamination (refer to Section~\ref{sss:41}). All luminosities are unabsorbed values scaled to a distance of 1.8 Mpc.
	\label{f:lightcurve}}
\end{figure}

We plotted the long-term $0.5-10$ keV light curves for the three ULXs  (sources 19, 25, 38) and one bright source (source 17) in Figure~\ref{f:lightcurve}. A distance to IC342 of 1.8 Mpc was assumed and the luminosities were not corrected for absorption along the line of sight. The baselines of the light curves were extended by adding data points from the 1998 \rosat\ HRI results and the 2006 \chandra\ HRC-I observations. These data were extracted from the archive and were reanalyzed using standard pipeline procedures of the corresponding instruments. The derived luminosities in this work cannot be compared directly with the previously published ones because of the different assumptions made in those works. For example, luminosity values in the earlier \asca\ and \rosat\ publications were computed assuming a distance to IC 342 of 4.5~Mpc instead of the 1.8~Mpc assumed here. We recalculated luminosities of all sources from all  observations by assuming the same energy range, spectral model, and assumed distance of IC 342. Identical extraction aperture for each source had been used in the \xmm, \rosat, and \chandra\ data, which was reliable for source 17, 19, and 25. However, with source 38 located in the crowded region near the center of IC 342, it suffers from contamination of nearby sources as well as the diffuse background of the center. Using a $r=15''$ circular aperture centered on source 38, it was estimated that roughly 40\% of source counts were contaminated by the faint source C13 \citep{Mak2007} which was unresolved in our \xmm\ images, and the local background when compared to the photometry obtained by~\citet{Mak2007}. This suggests that the flux values estimated for this source from \xmm\ and \rosat\ data are overestimated due to their larger PSF and as a result we should keep that in mind while examining its light curve.

The flux values plotted in Figure~\ref{f:lightcurve} were obtained from spectral fits to the four observations using XSPEC. For consistency, the fluxes of the four \xmm\ observations of each sources were derived from the same spectral model with identical best-fitting parameters (Table~\ref{t:spe}). On the other hand, with no spectral information in the \rosat\ HRI and \chandra\ HRC-I observations, the fluxes were instead estimated using PIMMS using spectral parameters averaged from the four \xmm\ observations. With the exception of source 38 which is known to be a spectral variable (see Section~\ref{ss:spe}), the spectral parameters of sources 17, 19, and 25 do not vary much among the four observations and thus the uncertainties in the derived \chandra\ and \rosat\ fluxes from these parameters for each source is small or negligible. For source 38, the \chandra\ and \rosat\ fluxes derived from the four sets of spectral parameters (using the power-law + blackbody + gaussian model) vary by a factor of $\approx1.3$. The implied flux ranges were accounted for in the error bars while the average value was plotted as the data point of its long-term light curve. For the other three sources in Figure~\ref{f:lightcurve}, the error bars in the light curves for \chandra\ and \rosat\ data points are dominated by  poisson errors in the counts, while those in \xmm\ data points are $1\sigma$ gaussian confidence limits from XSPEC. It should also be noted that comparisons of luminosities between different instruments should be treated with caution because various factors (e.g. cross-calibration issues and uncertainties in instrument responses and in assumed spectral shapes for energy conversions) could possibly contribute to the errors. Previous studies indicated that variations of normalized flux in different X-ray instruments were small, at a level of $\pm10\%$ \citep{Snowden2002}, and thus were not expected to account for all the flux variability observed. 

\subsection{X-ray Spectral Analysis and Spectral Variability}
\label{ss:spe}
We searched for spectral variability of our \xmm\ sources using two methods. We first inspected changes in the best-fit spectral shapes and/or measured changes in the best-fit parameters. Similar measurements of spectral variability had been used by~\citet{Grimm2007} to search for spectra variables in M33. However, this could only be done for bright sources that possessed enough source counts for reliable spectral fits. Second, we adopted a spectral variability factor analogous to the flux variability factor described in Section~\ref{ss:var}, with flux in equation~(\ref{eq:f}) replaced by the hardness ratios. A source is defined to be a spectral variable if $S_{\rm spectral}>3$. This method could be performed on all sources, including the faintest ones.

\subsubsection{Spectral Fitting}
We extracted the energy spectra of the 11 brightest sources from all our four \xmm\ catalog with the SAS task XMMSELECT. These 11 sources included those that were highly variable in flux or spectral. Response matrices were created by RMFGEN and ARFGEN while $\chi^{2}$ statistics was used to find the best-fitting parameters. We first tried simple absorbed one-component models including power-law, blackbody, multicolor disk black-body (diskbb model in XSPEC;~\citealt{Mitsuda1984}), Raymond-Smith (RS), BMC\footnote{The typical scenario involved thermal X-rays from the inner region of an accretion disk in a black-hole binary illuminating in-falling matter in close proximity to the black-hole event horizon. (http://heasarc.nasa.gov/xanadu/xspec/manual/XSmodelBmc.html)}, and broken power-law models. For a few cases two-component models were attempted when single-component model did not generate satisfactory fits. Spectra from the three \xmm\ detectors were fitted simultaneously for each source, allowing only an overall normalization factor to vary among detectors to account for differences in calibration. 

The spectra of these 11 sources were satisfactorily fitted with simple absorbed power-law models, except for source 27 (SNR candidate), 36, and 38 (the nuclear source). The best fit spectral parameters are listed in Table~\ref{t:spe}. The best fit $N_{\rm H}$ ranges from $2.3\times10^{21}\rm cm^{-2}$ to $2.5\times10^{22}\rm cm^{-2}$, with an average value of $8.0\times10^{21}\rm cm^{-2}$, while the photon index varies between 0.47 and 3.52. 
In general, at least 1000 detected counts were needed to clearly rule out competing spectral models.
In spite of this, the deduced fluxes do not vary significantly among different spectral models and thus it is possible to derive X-ray luminosities and study long-term flux variabilities. Detailed fitting result of three brightest X-ray sources (sources 19, 25, 38) and the SNR candidate (source 27) will be discussed below in detail in section~\ref{ss:ind}.

Five sources (19, 23, 24, 25, 38) were classified to be spectral variables based on spectral fits. For sources 19 and 25, their spectra shapes changed significantly not only among the four \xmm\ observations, but also over the past 10 years when compared to previous \asca\ observations. The changes are reminiscent of spectral transitions of black hole binaries (BHBs) which usually vary between the high/soft and the low/hard state. For source 38, the spectrum in 2001 was significantly different from the other three with a larger photon index at energy above 1~keV. We shall discuss in detail the X-ray properties of these three sources in Section~\ref{ss:ind}. The fitted spectral parameters of these five sources all showed variability at the 90\% confidence level  (corresponding to a 2.706$\sigma$ confidence limit). For example, the power-law fits of source 24 for the 2004 February and 2004 August data resulted in a change of photon index $\Delta\Gamma = 2.4$, implying a photon index change at more than 99\% confidence level (corresponding to a 3$\sigma$ confidence limit). The power-law + MEKAL fits to data of source 38 also showed a trend of decreasing photon index ($\Delta\Gamma = 1.3$) and increasing plasma temperature ($\Delta kT\approx0.37$) at the 90\% confidence level. 

\subsubsection{Spectral Variability Factor $S_{\rm spectral}$}
We define two spectral variability factors $S_{\rm spectral}$ by replacing the flux with HR1 and HR2 in equation~(\ref{eq:f}). High spectral variability with $S_{\rm spectral}>3$ in either HR1 and HR2 was detected only in three of the sources (source 19, 36, 41) in which two (19 and 36) had already been classified as spectral variables from spectral fits. 
An alternative method to search for spectral variable is to study the color-luminosity diagrams. We inspected the color-luminosity diagrams of all sources and found that 15 of them exhibited significant spectral transitions in different patterns.
Five of them (source 20, 32, 40, 45, and 51) showed typical spectral behavior reminiscent of Galactic X-ray BHBs which were seen as having negative slopes in the plots; one of them (source 52) showed correlation between lower luminosity and decreasing hardness; eight of them (source 9, 16, 27, 30, 34, 40, 43, 55, 56) showed more complicated behavior or underwent several transitions; one of them (source 38) varied spectrally but not in intensity. As a result, a total of 19 sources, or $\approx30\%$, were classified as spectral variables under these criteria and they were marked with "s" in Table~\ref{t:cat}. Spectral variability seen in X-ray sources is common in external galaxies.~\citet{Zezas2006} found 21 out of 120 X-ray sources, $\approx18\%$, in the Antennae show signs of variability in hardness ratios while~\citet{Fridriksson2008} found that 2 out of 90 and 3 out of 38 X-ray sources in NGC 6946 and NGC 4485/4490 respectively show significant ($3\sigma$)variability in the hardness ratios as defined similarly in this work. It is also noted that while these galaxies, just like IC342, are starburst galaxies, the fraction of spectral variables we identified  in IC342 is higher than these galaxies.

\subsection{Identifications of X-ray Sources}
\label{ss:ide}
Using properties of the detected sources (e.g. HR, variability) and cross-correlating them with catalogs at other wavelengths, we attempted to classify all sources in our X-ray catalog into different classes of X-ray emitting objects. We varied the searching radius according to the accuracy of the various catalogs and visual inspection was performed to confirm the spatial coincidence of a counterpart. 
A summary of our classification scheme is presented in Table~\ref{t:sch} and the results of the source identification are listed in Table~\ref{t:ide}. We briefly explain the classification of each type of objects below.

\subsubsection{Supersoft sources and Quasisoft sources}
\label{sss:sss}
Supersoft X-ray sources (SSS) are characterized by their soft X-ray emissions, which can be represented by blackbody spectrum with $kT\le100$ eV, and with X-ray luminosities of $10^{36-38}\lum$ \citep{Stefano2003a}. These sources show little or no X-ray emission above 1--2 keV, and are generally believed to be binary systems with white dwarfs accreting from more massive hydrogen burning donors \citep{Heuvel1992}. Some of them show significant variabilities on various time scales, while some other have been found to be recurrent transients~\citep{Greiner2004,Stefano2004,KongStefano2003,Osborne2001}. Quasisoft X-ray sources (QSSs) are luminous ($L_{\rm x} > 10^{36}\lum$, kT between 120 eV and 350 eV) X-ray sources emitting few or no photons at energy above 2 keV yet clearly emitting at above 1.1 keV \citep*{Stefano2003b,StefanoKong2004}. They also suggested that if we observe a hot SSS located behind a large gas column, just like the case of Galactic center, photons in the medium energy band (1.1--2 keV) would be detected, while only few soft photons would be detected. These sources might be the hottest nuclear burning white dwarf binaries and could possibly be progenitors of Type Ia supernovae, and SNRs.

Since most sources in our catalog had too few photons (counts $< 200$) for meaningful spectral fitting, we employed the selection algorithm for SSS and QSS defined by \citet*[fig.3]{Stefano2003b} based mainly on hardness ratios. The energy bands defined in the hardness ratios of their work were different from what we used earlier in Table~\ref{t:sch}, in which the three energy bands were defined as: soft ($S$) $=0.2-1.1$ keV, medium ($M$) $=1.1-2.0$ keV, and hard ($H$) $=2.0-7.0$ keV. We therefore recalculated hardness ratios according to these definitions. The summary of source counts and hardness ratios of the identified SSS and QSS is given in Table~\ref{t:sss}.

Excluding those that had been classified as foreground stars or candidates, we identified source 36 and source 56 as confirmed SSS and QSS respectively, and we describe source 36 immediately below in detail. Source 56 satisfied the QSS selection criteria in the 2002 February data but was too faint to be identified as a source in the remaining three observations. Two additional sources (52, 54) also satisfied the QSS selection criteria. Source 52 satisfied the selection criteria in 2005 February in which the S/N is the lowest among the four observations. It however possessed substantial hard signal in the other three observations. Source 54 was identified based on the 2004 February and 2005 February observations in which the X-ray emission are dominated by photons below 2~keV. On the other hand, substantial hard signal was present when the source was in its high state in 2004 August. We therefore classified both sources 52 and 54 as possible QSS candidates and also presented their properties in Table~\ref{t:sss}. Interestingly, all the classified QSS and QSS candidates are under the category of "QSS-noh", meaning little or no hard X-ray emission was detected. 

The only classified SSS, source 36, had also been identified as a transient. It was very luminous during its high state in both observations in 2004 but disappeared from the field in the 2005 February observation, just after it reached the highest state in 2004 August. Spectra of the source were extracted from the two 2004 observations. Simple models all gave unacceptable fits with $\chi^{2}\geq2$. Satisfactory fits could be obtained only after a gaussian line at $\approx0.9$ keV was added to the blackbody and power-law model. A relatively high temperature of $kT_{\rm bb}\approx0.17$ keV was obtained for the BB model, while very large photon indices of $\Gamma=3.6$ and 6.8 were obtained for the power-law model with the 2004 February and August data respectively, indicating soft X-ray emissions. We noticed that SSS transients are common in nearby galaxies, with two luminous examples being source 110 in NGC4697 \citep{Sivakoff2008} and ULX-1 in M101 \citep{Kong2004b,Kong2005}. They both belong to the group of ultraluminous SSS (ULS) that have been suggested to be accreting intermediate mass black holes (IMBHs) \citep{Stefano2003a}. In our case, source 36 is likely to be consistent with the white dwarf model since its luminosity conforms to the near-Eddington luminosity of a 1.4$M_{\odot}$ white dwarf ($L_{\rm X}^{\rm Edd}\sim1.8\times10^{38}\lum)$. We looked for its optical counterpart from the Digitized Sky Survey (DSS) data and UV/optical counterpart from the \xmm\ Optical Monitor (OM) images, but found no source at its position in all images.  

\subsubsection{Supernova Remnants}
\label{sss:snr}
With IC342 not being an active target for SNR survey, there had been few identified SNR candidates in the galaxy. We searched for SNR and stellar novae as listed in the International Astronomical Union Circulars (IAUCs), and no matches within $10'$ of IC 342 was found. On the other hand, we found four SNR candidates listed in the optical search by~\citet{Dodorico1980}. The SNR object 1 (based on designation of~\citealt{Dodorico1980}) is $9.4''$ away from our X-ray source 27, with uncertainty in position of the SNR at about $10''$ from the optical data, while the other three SNR in~\citet{Dodorico1980} were not within the proximity of any of our \xmm\ sources. This SNR was first identified in radio continuum observations of the galaxy~\citep{Baker1977}, and was later confirmed as a SNR on the basis of optical spectroscopy~\citep{Dodorico1980}. Detailed X-ray analysis of this source will be presented in Section~\ref{sss:29}. Its X-ray spectrum could be fitted with an absorbed NEI model (c.f. Section~\ref{ss:ind}) with a derived 0.5--10 keV luminosity at $\sim 9\times10^{36}\lum$, which is similar to that of SNRs in the Magellanic Clouds (e.g.~\citealt*{Hughes1998},~\citealt{Williams1999}). Its X-ray spectrum, combined with its proximity to an optical SNR suggest that source 27 is a strong SNR candidate.

In addition to comparison with optical SNR catalog, we also attempted to identify SNR using the hardness ratio criteria outlined in~\citet{Misanovic2006}, which required a SNR candidate to have $HR1_{\rm z} > 0.1$ and $HR2_{\rm z} < -0.4$, where $HR1_{\rm z}=(S_{2}-S_{1})/(S_{2}+S_{1})$ and $HR2_{\rm z}=(S_{3}-S_{2})/(S_{3}+S_{2})$, with $S_{1}$, $S_{2}$, and $S_{3}$ the source counts in the energy bands $0.2-0.5\rm~keV$, $0.5-1.0\rm~keV$, and $1.0-2.0\rm~keV$ respectively. Excluding the foreground candidates (see below), two sources (48, 59) were found to satisfy this criteria. It was noted that the aforementioned SNR candidate source 27 did not satisfy the HR criteria. As we will discuss in Section~\ref{sss:29}, the X-ray spectrum of source 27 could also be satisfactorily fitted by a power-law model and thus it was possible that a hard component resided in its X-ray emission. Together with the fact that the hardness ratio criteria of  \citet{Misanovic2006} only considered X-ray emission below 2 keV, this source could therefore possibly be missed. It also implied that similar sources with a relatively hard component could be missed with this criteria as well. 

For the two SNR candidates selected by the HR criteria, we checked for their variabilities as outlined in Section~\ref{ss:var}. We found that source 59 exhibited some degrees of variability in flux, with $S_{\rm flux} = 4.7$.  Since the X-ray emission of a SNR is expected to be persistent, this source was rejected from being a SNR candidate. On the other hand, source 48 had a very soft X-ray emission which was consistent with a QSS classification. 
It had an optical counterpart in the USNOB1.0 catalog with an offset of $0.58''$. Moreover, the X-ray to optical flux ratio at $\log(f_{\rm x}/f_{\rm opt}) = -0.3$ was larger than the expected value for foreground stars, which should have $\log(f_{\rm x}/f_{\rm opt})<-1$ \citep{Maccacaro1988}. Since optical emission of accreting binary systems at the distance of IC342 is below the sensitivity of the USNO and 2MASS catalogs, therefore source 48 was classified to be a SNR.  

\subsubsection{Foreground Stars}
\label{sss:for}
We compared our \xmm\ source list with the USNO~B1.0 catalog and found that 25 of them had optical counterparts within a $5''$ searching radius (except for source 59 which had an offset of $6.2''$ with an USNO star, but the searching radius was relaxed because it was at a large off-axis position). We found 13 of these 25 sources have X-ray to optical flux ratios that were consistent with those of normal stars of $\log(f_{\rm x}/f_{\rm opt})\leq1$, using the criteria of~\citet{Maccacaro1988}. We calculated the X-ray to optical flux ratios from $\log(f_{\rm x}/f_{\rm opt})=\log(f_{\rm X})+0.4V+5.37$ for each source. In the calculations, the X-ray flux $f_{\rm x}$ was calculated by assuming a simple power-law model restricted in the energy range  0.3--3.5 keV with $\Gamma=2$ and $N_{\rm H}=8\times10^{21}\rm cm^{-2}$ (same model as assumed in luminosity values in Table~\ref{t:cat}), and the value of the {\it V} magnitude was averaged from the {\it B} and {\it R} magnitude in the USNO~B1.0 catalog. To further constrain the classification of these 13 sources as foreground stars, two additional criteria are proposed:
\begin{enumerate}
\item Hardness Ratio: X-ray emission from stars are relatively soft from studies of their spectra or hardness ratios. The energy spectrum of a foreground star could be best fitted by a Raymond-Smith model with $kT_{\rm RS}\approx1$ keV or a power-law model with a photon index $\Gamma\ge3$~\citep{Kong2002}. In our \xmm\ catalog, 13 sources were not bright enough for spectral analysis. Instead we used the hardness ratio criteria defined in~\citet{Misanovic2006}, which required a foreground stars to have $HR2_{\rm z}<0.3$ and $HR3_{\rm z}<-0.4$, where $HR3_{\rm z}=(S_{4}-S_{3})/(S_{4}+S_{3})$ and $S_{4}$ was the source counts in the energy band $2.0-4.5\rm~keV$, while $S_{3}$ and $HR2_{\rm z}$ were the same as that defined in Section~\ref{sss:snr}. 
\item 2MASS counterpart: For sources that had near-IR counterparts from the 2MASS catalog, we checked for their near-infrared colors and magnitude. Adopting the criteria of \citet{Finlator2000}, we classified sources with $J-K<0.8$ and $J<12.5$ as foreground stars.
\end{enumerate}

To classify sources as foreground stars, four criteria had been proposed in which the first two, that is, USNO counterparts and X-ray to optical flux ratios, were used for screening purpose. Based on the two additional criteria, we classified X-ray source to be a foreground star of category 1 ("Cat1" in Table~\ref{t:ide}) if all four criteria were fulfilled. Sources that satisfied either the hardness ratio or the 2MASS criteria in addition to the two screening criteria were identified as foreground star candidates and were marked as categories 2 and 3 respectively ("Cat2" and "Cat3" in Table~\ref{t:ide}). We confirmed four sources (7, 24, 26, 29) as category 1 foreground stars that satisfied all the criteria described above. These four sources were also identified as foreground stars by~\citet{Bauer2003} based solely on the X-ray to optical flux ratio. Furthermore, we identified seven other sources for the first time as foreground star candidates since they satisfy either the hardness or the 2MASS criteria. The majority of these candidates, six in total (source 16, 28, 37, 43, 51, 61), belonged to the category 2, with only one (source 2) belonging to category 3. 

In summary, a total of 11 sources were classified as foreground stars or candidates. They all had bright optical counterparts, with visual magnitude $B < 20.2$ and $R < 17.1$. Except for sources 2, 16, and 61, they all showed significant X-ray variability on the time scale of years, probably due to flarings. In particular, sources 43 and 51 were transients and exhibited  large changes in flux. Source 24 was the only star candidate that had enough counts for reliable spectral fitting. Both the Raymond-Smith and power-law models could generate satisfactory fits to the spectrum, with $kT_{\rm RS}\sim0.9$ and power-law index $> 3$ (c.f. Table~\ref{t:spe}), which was consistent with the spectrum of foreground stars.


\section{PROPERTIES OF X-RAY SOURCES}
\label{s:pro}
\subsection{Global Properties}
\label{ss:glo}
Analysis of the four observations of IC 342 with \xmm\ between 2001 February and 2005 February were presented in previous sections. The limiting luminosity of these exposures was between $\sim(1-2)\times10^{37}\lum$ depending on the local background and exposure time. 
Combining all four observations, a total of \srcno\ X-ray sources were detected. While most detected sources were too faint for detailed analysis, we used the X-ray colors, flux and spectral variability, and in some cases optical or radio counterparts to classify the detected X-ray sources into several stellar populations.  We identified a total of 11 foreground stars, 1 SSS, 3 QSS (one confirmed and two candidates), and 2 SNRs, with a summary given in Table~\ref{t:ide}. The spatial distribution of all detected sources was shown in Figure~\ref{f:ide}, with classified sources characterized in different symbols. The SSS/QSS were found to be located near or on the spiral arms, associating them with young stellar populations. On the other hand, the two SNR are very close to the starburst nucleus dominated by current star formation activities. 

One major focus of this work is the study of intensity and spectral variability of the X-ray sources in IC342 on timescales of years. We found that 41 of the \srcno\ detected sources, or 64\%, showed long-term flux variability, clearly indicating that they were individual X-ray binaries. Of these, 26 sources, or $\approx43\%$, were classified as X-ray transients. The observed fraction of sources showing flux variability of (64\%) is quite high compared to previous studies of late-type galaxies. For example, 27\% of the X-ray sources in the 11 nearby face-on spiral galaxies studied by~\citet{Kilgard2005} exhibited variability on either long or short timescales, 25\% of the sources in M33 exhibited long-term variability~\citep{Grimm2007}, 29\% of the sources in NGC6946, and 39\% of the sources in NGC4485/4490 were variables on timescales from weeks to years~\citep{Fridriksson2008}. This indicated that the source population in IC342 was dominated by accreting XRBs, in agreement with the results of~\citet{Kong2003} that sources in IC342 were mostly HMXBs. In addition to flux variability, 19 sources, or 30\% of total, were identified to be varying in hardness ratios or undergoing spectral transitions. Eight of the identified X-ray transients showed spectral variability in addition to flux variability as seen from the color-luminosity diagrams. 

\begin{figure}[htbp]
\begin{center}
\leavevmode	
	\includegraphics[width=14cm, angle=0]{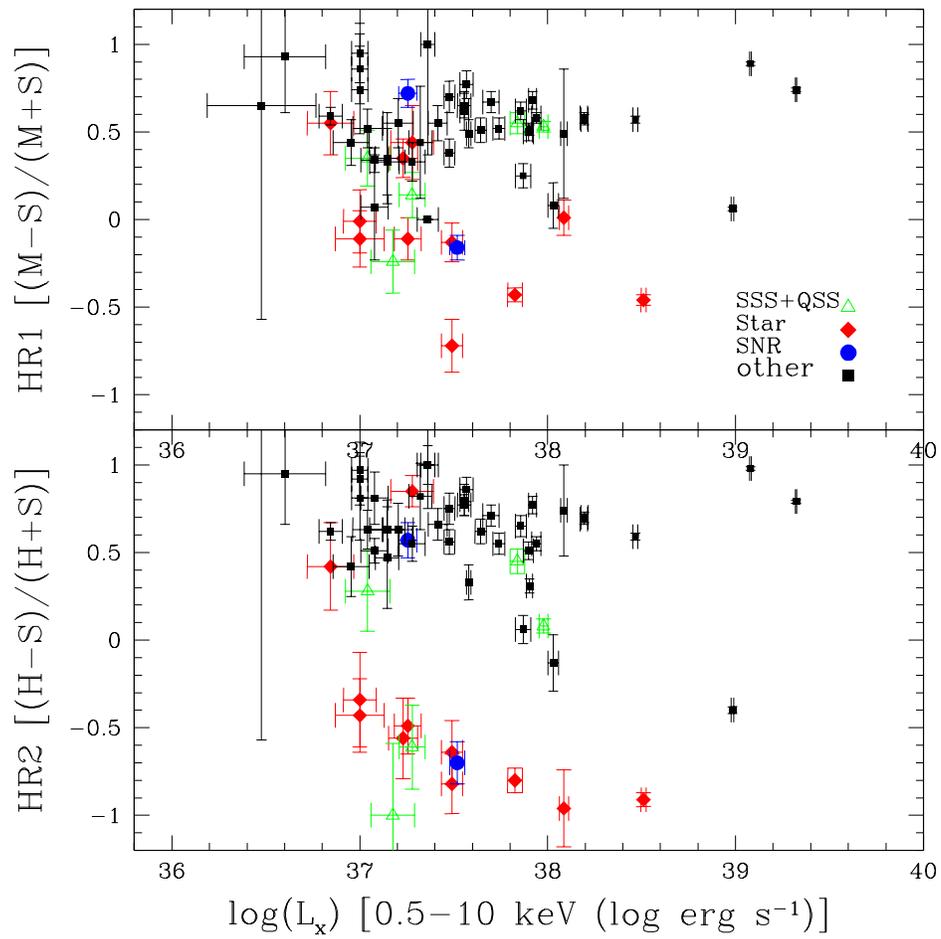}\\
\end{center}
	\caption{Hardness-intensity diagram for all sources detected in the four \xmm\ observations. All HR1, HR2, and luminosity are average values as given in Table~\ref{t:cat}. 
	\label{f:har}}
\end{figure}

In Figure~\ref{f:har}, the X-ray hardness ratios (HR1 and HR2) versus the 0.5--10 keV  X-ray luminosities of all detected sources were plotted. Most sources were in the luminosity range $10^{37}-10^{38}\lum$, and their hardness spanned a diverse population. The HR1 color-luminosity diagram (upper panel) showed that a majority of the sources are hard while the HR2 diagram (lower panel) revealed a large proportion of hard sources along with a separate and smaller soft sources. This soft population in HR2 color comprised of foreground stars, SNRs, SSS/QSS, 
with only one source that was unclassified. This is consistent with the expectation that these sources are characterized by their soft color. As pointed out in~\citet{Zezas2006}, the observed HR1 color was sensitive to absorption while HR2 was most sensitive to the intrinsic spectral shape. With the X-ray sources in IC342 being mostly hard and slightly obscured, this picture is consistent with a population dominated by HMXBs, as would be expected in sites of recent star formation. For a comparison with an extragalactic X-ray source population, \citet*{Irwin2003} studied 15 nearby early-type galaxies observed with \chandra\ and found that sources with luminosities in the $(1-2)\times10^{39}\lum$ range had softer spectra (power-law $\Gamma\sim2$), which was consistent with the high/soft state of black hole binaries. On the other hand, the two sources in IC342 with luminosities above $10^{39}\lum$ were instead generally harder, with hardness ratios $> 0.5$. 

\subsection{Individual Sources}
\label{ss:ind}
\subsubsection{Source 19 (IC342 X-1): An ultraluminous compact X-ray source}
\label{sss:21}
Source 19 (IC342 X-1) is the most studied X-ray source in IC342. It was discovered by \einstein~\citep{Fabbiano1987} and confirmed as a point source with \rosat\ observations~\citep{Bregman1993}. Subsequent \asca\ observation in 1993 September showed that this source was in a very luminous state with absorbed $L_{\rm 0.5-10\rm~keV}=3.8\times10^{39}\lum$~\citep{Kubota2002} (scaled to distance 1.8 Mpc) and its X-ray spectrum was best-fitted by an absorbed multi-color disk model (disk blackbody model in XSPEC), with $N_{\rm H}=(4.7\pm0.3)\times10^{21}\rm cm^{-2}$ and $T_{\rm in}=1.77\pm0.05~\rm keV$, which is generally used to describe ULX spectra~\citep{Okada1998}. This corresponded to  the characteristics of a black hole accretion disk in the high/soft state. The source dimmed by a factor of three in a follow-up \asca\ observation in 2000 February with absorbed  $L_{\rm 0.5-10\rm~keV}=1.2\times10^{39}\lum$ (scaled to distance 1.8 Mpc) and the spectrum changed dramatically to an absorbed power-law model, with $N_{\rm H}=(6.4\pm0.7)\times10^{21}\rm cm^{-2}$ and $\Gamma=1.73\pm0.06$~\citep{Kubota2002}. The power-law-spectra as seen in many ULXs can be explained by the low/hard state that are observed in many Galactic and Magellanic BHB systems.~\citet{Kubota2002} noticed that there was significant softening in the X-ray spectrum above 5 keV and instead proposed that it was associated with an anomalous very high (VH) state (or recently described as steep power-law (SPL) state by~\citealt{Remillard2006}), also seen in many Galactic black hole binaries (e.g. GX 339-4,~\citealt{Markoff2003}; GRS 1915+105,~\citealt{Belloni2000}). The VH state is characterized by strong Comptonization and the \asca\ spectra in 2000 could indeed be adequately fitted by a strongly Comptonized optically thick accretion disk with $T_{\rm in}=1.1\pm0.3~\rm keV$ and the exponent of the radial dependence of the disk temperature $\Gamma_{\rm th}=2.2\pm0.4$.

\begin{figure}
\centering
\begin{tabular}{cc}
\epsfig{file=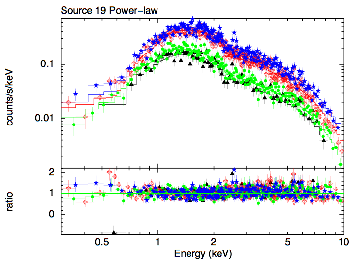, angle=270,width=0.5\linewidth,clip=} & 
\epsfig{file=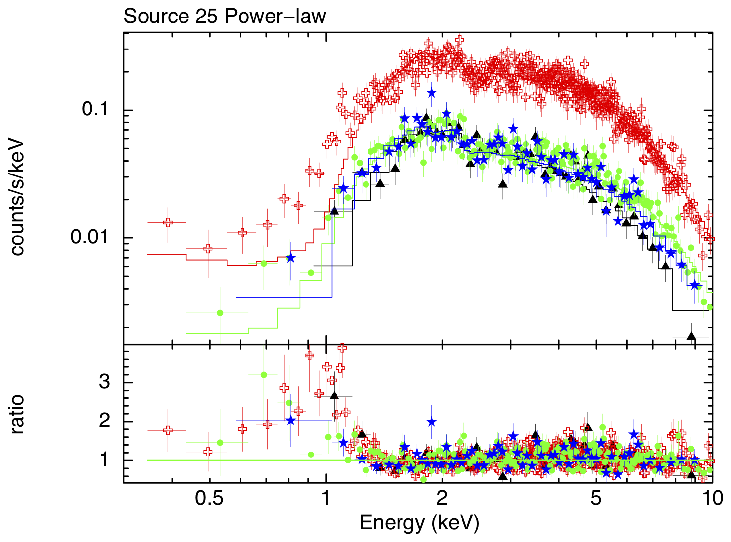, angle=270,width=0.5\linewidth,clip=} \\
\epsfig{file=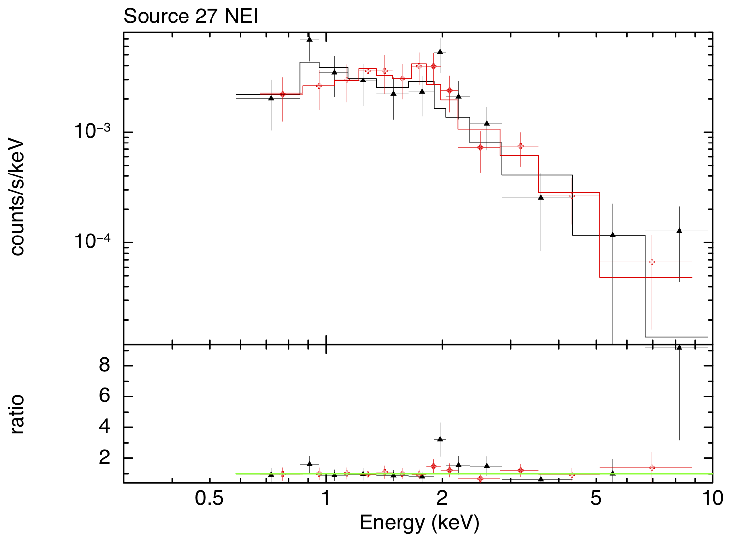, angle=270,width=0.5\linewidth,clip=} &
\epsfig{file=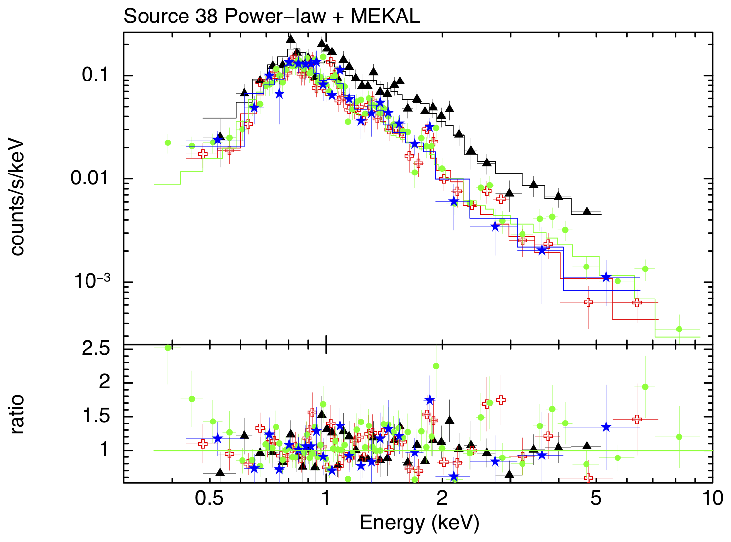, angle=270,width=0.5\linewidth,clip=} 

\end{tabular}
\caption{\xmm\ EPIC-PN energy spectra of the three brightest X-ray sources: source 19 (IC342 X-1; top-left), source 25 (IC342 X-2; top-right), and source 38 (IC342 X-3 nuclear source; bottom-right). The fourth plot is the source 27 (SNR candidate; bottom-left). The 2001 February (black triangle), 2004 February (red cross), 2004 August (green circle), and 2005 February (blue star) data are shown together with the lines showing the best-fit spectra. The best-fit parameters are given in Table~\ref{t:spe}. 
\label{f:spectrum}}
\end{figure}

Using the \xmm\ data from 2001 February to 2005 February, the 0.3--10 keV spectra of source 19 were fit with absorbed power-law models (Figure~\ref{f:spectrum}, top-left) with spectra parameters listed in Table~\ref{t:spe}. The model gave satisfactory fits to all these observations except for the longest exposure 2004 August data. The photon index and the absorption column density increased from 2001 February to 2004 February, followed by a slight drop six month later in 2004 August, while the inner disk temperature showed the opposite transition during the same period. It is worth to note that the spectral parameters of the disk blackbody model fit in the 2005 February data are remarkably similar to that of the 1993 \asca\ observation. The spectral parameters from the power-law model are also consistent with those obtained from the 2000 \asca\ observation. However, if we restricted the analysis to the energy range 2--10 keV, the power-law fits to the data showed that only the photon index in the 2001 February data was steep enough ($\Gamma\approx2.4$), to be marginally consistent with the spectrum of the VH state as defined by the presence of power-law component with $\Gamma\geq2.4$ in the 2--20 keV band~\citep{Remillard2006}. The photon indices of the subsequent three observations were all lower than 2 in the 2--10 keV band. 
For the X-ray luminosity, the values in the energy range 0.5--10 keV also changed dramatically over the $\sim$15 years. The long-term lightcurve of IC342 source 19 in Figure~\ref{f:lightcurve} (top-right) showed that it was one of the most variable sources in IC342 and its luminosity increased by a factor of more than 3 from 1991 to 2006. We noted that (and also pointed out by~\citealt{Kong2003} and~\citealt{Bauer2003}) the \asca\ observations  
suffered from serious confusion problem. The large extraction radius ($3'$) used in analyzing source 19~\citep{Okada1998} would have included nine other sources (source 17, 18, 21, 24, 26, 27, 28, 29, and 30) detected in our four \xmm\ observations. Using the present \xmm\ photometry, about 24\% of the \asca\ counts of source 19 could be due to confusion. We therefore excluded the \asca\ data of 1993 and 2000 from the lightcurve shown in Figure~\ref{f:lightcurve}. On the other hand, it was noted that the luminosity during the high state in 1993 as suggested by the \asca\ data, after subtraction of  the flux contribution from confusion, at $L_X\sim2.9\times10^{39}\lum$ would be similar in magnitude to the high-level \chandra\ flux in 2006 with $L_X\sim3.1\times10^{39}\lum$. 

Based on the results above, we proposed that source 19 had undergone multiple state transitions. Starting from a low state in the 1991 \rosat\ observation, it changed to the high/soft state in the \asca\ observation in 1993, changed again to the VH state in the follow-up \asca\ observation in 2000 and remained in this state until the first \xmm\ observation in 2001 February. It then returned to the high/soft state in 2004 February and changed to the low/hard state six months later in 2004 August. In 2005 February, its luminosity was more than double the intensity level of 1993 and stayed at the high/soft state through to the 2006 \chandra\ observation. If source 19 was truly in the VH state during the \asca\ observation in 2000 and the \xmm\ observation in 2001, it would be the longest period that a VH state was observed~\citep{Bauer2003}. Finally, we searched for short-term variability within each \xmm\ data but found none. 

The spectral/intensity variability of source 19 suggests this source is most likely a compact accreting object of a black-hole binary. A plausible scenario for the ULX source 19 is that it is an IMBH formed via mergers of massive stars/BH in a compact star cluster, as supported by the Comptonized disk model spectrum~\citep{Kubota2002}. On the other hand, we cannot rule out other possible explanations for the super-Eddington luminosity, such as a stellar mass black hole with strongly beamed X-ray emission~\citep{King2001,Kording2002}. Moreover, source 19 was associated with a "tooth"-shaped optical nebula~\citep{Pakull2002}.~\citet{Roberts2003} found two regions of \ion{O}{3} emission located on the inside of the shell of the nebula and suggested they were caused by photonization of the nebula shock excited from the ULX source 19. This scenario was later confirmed by~\citet{Abolmasov2007}.~\citet{Feng2008} suggested that the incomplete shell in optical morphology could be a jet emission if the nebula was powered by an unusual powerful explosion in which a black hole was formed in source 19.

\subsubsection{Source 25 (IC342 X-2): An ultraluminous compact X-ray source}
\label{sss:27}
This source is the second brightest ULX in IC 342 discovered by the \einstein \citep{Fabbiano1987}. The spectrum taken by \asca\ in 1993 showed that it was in the low state, with absorbed $L_{\rm 0.5-10~keV}\approx1.6\times10^{39}\lum$ scaled to 1.8 Mpc~\citep{Kubota2001}, and could be equally satisfactorily fitted by both power-law and disk blackbody model. The best-fit absorbed power-law parameters were $N_{\rm H}=(14.3\pm1.6)\times10^{21}\rm cm^{-2}$ and $\Gamma=1.39\pm0.10$~\citep{Okada1998}. The second \asca\ observation in 2000, with absorbed $L_{\rm 0.5-10~keV}\approx2.8\times10^{39}\lum$ scaled to 1.8 Mpc, revealed that the spectrum of source 25 was more convex than in 1993 and could be expressed with a disk blackbody model of $N_{\rm H}=(18.2\pm0.8)\times10^{21}\rm cm^{-2}$ and $T_{\rm in}=1.62\pm0.04$ keV, whereas the power-law fit was unacceptable~\citep{Kubota2001}. The flux increased by a factor of $\sim2$ between these two observations. Such a transition from low/hard state to high/soft supported the black hole interpretation of source 25. The four \xmm\ spectra of source 25 could be satisfactorily fitted by both the power-law and disk blackbody (diskbb) models, with the diskbb model giving consistently better fits than power-law models (Figure~\ref{f:spectrum}, top-right). This was consistent with the spectral fits of the 1993 \asca\ data for the source. The spectra had soft excess below 1 keV and was the flattest in 2004 August with $\Gamma = 1.3$. In addition, the spectrum in 2004 August was different from the other three in which the inner disk temperature ($T_{\rm in}$) from diskbb model was particularly high together with a low photon index. Therefore, we also attempted the combined power-law and diskbb model and found that it gave a better fit to the soft excess seen below 1 keV with $T_{\rm in}=1.33\pm0.14$ keV and $\Gamma=1.15\pm0.18$. Similarly, a combined model of power-law and mekal gave similar satisfactory fit with $kT=0.16\pm0.05$ keV and $\Gamma=1.54\pm0.05$. Both models suggest the source was soft in 2004 August. These observations indicated that the source was undergoing significant spectral change from 2001 to 2005. 

Source 25 is the most variable source in our \xmm\ catalog with a variability factor $S_{\rm flux}\approx60$. It became very luminous in the 2004 February data (unabsorbed $L_{\rm 0.5-10~keV}\approx4\times10^{39}\lum$, even brighter than source 19) and in the 2006 \chandra\ observation (unabsorbed $L_{0.5-10\rm~keV}\approx3\times10^{39}\lum$), implying an increase of a factor of $> 4$ in flux from its faintest state. With the low state being caught (2004 August and 2005 February) in between the recorded high states, the timescale of the high/soft-low/hard transition is much shorter for source 25 than for source 19. The intrinsic spectral/intensity change of source 25 is high enough to explain the transition between the 1993 and 2000 \asca\ observations and thus the contamination of nearby sources did not notably affect the observed variability if comparable magnitude of spectral/intensity transitions also occurred in the \asca\ data. Besides long-term variability, source 25 also exhibited short-term variability with a periodicity of 31 hr or 41 hr found in a long ($\sim250$ ksec net exposure time) \asca\ observation \citep{Sugiho2001}.  

Similar to source 19, source 25 also suffered from the confusion problem with a large extraction radius of $3'$ being used. In our \xmm\ images, sources 23 and 34 in the \asca\ analysis were included in the $\rm r=3'$ circle centered on source 25, implying that the flux measured in the \asca\ observations could be over-estimated by $\sim10\%$. However,~\citet{Kong2003} noted that during the high/soft state in 2000, the source 25 was asymmetrically extending towards the direction of source 36, which was identified as a SSS located $\sim 3'$ away. This suggested that the flux estimates of source 25 could also be affected by source 36. Source 36 is highly variable with flux changing by a factor of 5 between the observed faintest and brightest state and its brightness was similar to that of source 25 in 2004 August. In addition, the asymmetry of source 25 in 2000 might imply that source 36 was in its bright state, thus contributing more soft photons to result in a softer and brighter spectrum in source 25~\citep{Kong2003} as well as the spectral variability observed during 1993 and 2000.  Assuming nearby sources contributed also to the flux of source 25 in the \asca\ observations in 1993 and 2000, the source could be dimmed by more than a factor of 2 from its high state in 2000 (with unabsorbed $L_{\rm 0.5-10~keV}\approx2\times10^{39}\lum$) to low state in 2001 (with unabsorbed $L_{\rm 0.5-10 \rm~keV}\approx1\times10^{39}\lum$). All these supported the proposition that source 25 is an accreting binary object. Nevertheless, there is a lack of multiwavelength analysis for this source. Studies of optical or radio counterparts will help to identify the nature of the compact object.

\subsubsection{Source 38 (IC342 X-3): The nuclear X-ray source}
\label{sss:41}
Source 38 is one of the three historical ULXs detected by \einstein. Recent high resolution X-ray imaging studies showed that the source was associated with the galactic center and thus confirmed it was not a ULX~\citep{Mak2007}. A comprehensive study of the multiwavelength and spatial analysis of this source based on the \chandra\ HRC-I observation in 2006 was also presented in that paper. Here we focus on the spectral analysis based on the four \xmm\ observations.

We found that simple models such as power-law, MEKAL, Raymond-Smith, and MCD all gave unacceptable fits, with $\chi^{2}_{\upsilon}\geq1.7$.~\citet{Bauer2003} obtained a good fit for source 38 using the \xmm\ data in 2001 with a best fit model of an absorbed MEKAL + power-law model ($kT=0.30^{+0.33}_{-0.07}$ keV, $\Gamma=2.52^{+0.15}_{-0.18}$, $N_{\rm H}=6.4^{+0.7}_{-1.0}$). We therefore attempted to fit source 38 with this model and with a combination of different absorption models. We found a combination of MEKAL and power-law model with Galactic absorption gave the best fit. Parameters of the spectra fits were listed in Table~\ref{t:spe}. The best fit $N_{\rm H}$ was about $(3.5-7.2)\times10^{21}\rm cm^{-2}$, similar to the result of \citet{Bauer2003}. The best-fit photon index $\Gamma$ showed a general decreasing trend from 2001 February (2.6) to 2005 February (1.5). Except for the 2004 February data ($\Gamma =2.2$), the spectra of source 38 were outside the typical range for AGNs ($\Gamma =1.7-2.3$).  In addition, there were also trends of decreasing absorption column density, and increasing plasma temperature ($kT$) from 2001 to 2005, strongly supporting a true spectral change. 

Closer inspections of the spectral shapes over the 4 years of \xmm\ data also revealed that the spectrum in 2001 was very different from the others (Figure~\ref{f:spectrum} bottom-right). There was an abrupt change in the slope for emission above 1 keV which showed a much harder emission in 2001 than in the follow-up observations. On the other hand, emission lines at $\sim$0.8 keV (Fe L), 1.1 keV (Ne Ly$\Gamma$), 1.4 keV (\ion{Mg}{12}), and 1.9 keV (\ion{Si}{13}) were most prominent in the spectra of both 2004 observations, primarily because of the higher signal-to-noise of these data sets. These emission lines were also observed in the nucleus of NGC 1808~\citep[see][fig.~6]{Jimenez2005}, and we adopted their line identifications here. No Fe K$\Gamma$ line was detected with high significance. This motivated us to replace the MEKAL model with one in which the abundances of individual elements could be fitted, that is, the VMEKAL model. The abundances of O, Ne, Mg, Si, and Fe were left as independent free parameters in the fits while the abundances of other elements are set to solar values. The best-fit abundances were sub-solar, except for the 2004 February observation in which an abundance of $\geq18~Z_\odot$, much higher than the value obtained by B03 ($Z=6.39Z_\odot$ for MEKAL component) was found. 
The values of the best-fit $\chi^{2}_{\upsilon}$/dof of the VMEKAL and MEKAL models were compared independently in each observation using F-test. The F-statistics
suggested that the $\chi^{2}_{\upsilon}$ improvement obtained when replacing MEKAL with VMEKAL was statistically significant only for the two highest S/N 2004 observations. In addition, an intrinsic absorption to the power-law component did not improve the fit and this intrinsic column density $N_{\rm H}$ gave a value of zero. 

An alternative model, with a power-law + blackbody was also compatible with the data, and had been used to fit the spectrum for the 2001 observation by~\citet{Kong2003} ($kT=0.11$ keV, $N_{\rm H}=8.7\times10^{21}$cm$^{-2}$, $\Gamma=2$). Even after we revised the model by adding a gaussian line at $\sim0.8$ keV to provide best fits there were large deviation in the spectra fits through the whole spectrum, and the power-law + MEKAL offered better fits to the data. Still, this model gave a reasonable estimate to the luminosity of the source. 

A final remark is that the variation of the spectral fits of source 38 to both power-law and MEKAL models is very consistent. It is interesting to note that spectral change seen in nuclear X-ray source is very rare and the change seen in source 38 would be supplement to this rare sample. On the other hand, the underlying mechanism that drives this spectral change in galactic nuclear source is still unclear. Further investigations are needed to understand the emission mechanism.

\subsubsection{Source 27: A Supernova Remnant candidate}
\label{sss:29}
As described in Section~\ref{sss:snr}, source 27 had been identified as a supernova remnant associated with the SNR object 1 of \citealt{Dodorico1980}.  We could not identify any X-ray sources from previous observations with \einstein, \asca, and \rosat. It was first detected in the 2001 February \xmm\ observation and was detectable in all observations at $L_{\rm X}\approx1\times10^{37}\lum$ until the last observation in 2006 by \chandra\ (C7 in~\citealt{Mak2007}). 
While the detection sensitivity of \einstein\ and \asca\ were an order of magnitude worse than \xmm\ and \chandra, \rosat\ had comparable sensitivity as these two satellites and had observed the field of source 27 in 1991. We used PIMMS to estimate the~\rosat\ PSPC count rates at the source position. Assuming the source flux had not changed significantly over the years, and using a power-law model with $\Gamma=2$ and $N_{\rm H}=8\times10^{21}$~cm$^{-2}$, the deduced \rosat\ count rate for source 27 would have been $3.8\times10^{-4}$ cps in the $0.1-4$ keV band.
We compared this predicted count rate with the background subtracted source counts of the \rosat\ data at the source~27 position using DMEXTRACT. The measured \rosat\ count rate ($1.9\times10^{-4}$ cps) is found to be at least a factor of 2 below the expected. This could suggest a possible scenario of an X-ray binary associated with the SNR that turns the source off and on again (e.g.\citealt{Williams2005,Williams2007}). This could be confirmed with a long-term monitoring of the source in the future.
This SNR was described as a diffuse shell with an angular size of 42 pc (assumed distance of 2.9 Mpc) in the optical images~\citep{Dodorico1980}. After scaling to our assumed distance of 1.8 Mpc, this corresponds to a size of $\approx$16 pc. However, source 27 was not resolved in the high spatial resolution \chandra\ HRC-I data and was consistent with a point source when compared to the \chandra\ PSF of size $\theta\approx2''$ (corresponding to 18 pc). This X-ray spatial extent is comparable to that in the optical images. It is worth noting that there are only a few spatially resolved X-ray SNRs beyond the Milky Way and Magellanic Clouds observed recently (e.g.~\citealt{Kong2002b,KongSNR2003,Kong2004}), and the unresolved X-ray structure of source 27 is possibly due to its intrinsic size being smaller than the \chandra\ PSF.

We noted that the Raymond-Smith (RS) and nonequilibrium ionization (NEI) models were often applied to study X-ray emission from extragalactic SNRs (e.g.~\citealt{Kong2002b,Schlegel2000}). The RS model is a simple collisional equilibrium ionization model, while the NEI model is appropriate for modeling SNRs whose ages are smaller than the time required to reach ionization equilibrium. The NEI model consists of an electron temperature ($kT_{\rm e}$) and an ionization timescale ($n_{\rm e}t$), where $n_{\rm e}$ and {\it t} are the mean electron density and the elapsed time after the plasma is shock heated to a constant temperature $kT_{\rm e}$ respectively. Due to limited counts, we were only able to fit the spectra of the two 2004 observations. The NEI model gave much better fits to the data than other simple models (including RS) since these simple models gave only acceptable fits to the spectrum ($\chi^{2}_{\upsilon}>2$) in 2004 August.
Results of the spectral fits, assuming solar abundance, were given in Table~\ref{t:spe} and Figure~\ref{f:spectrum} (bottom-left).  We attempted to fit the abundances for the NEI model but could not constrain the parameters. Moreover, there are indications of line emissions at 0.9 and 2 keV.
It is likely that the 0.9 keV feature comes from the Fe L shell lines and Ne K shell lines. The fitted electron temperature at above 2 keV was relatively high but it had been also observed in some extragalactic SNRs (e.g., N132D in LMC~\citealt{Favata1997}) which could be due to a shock-heated swept-up circumstellar medium or a inhomogeneity of the interstellar medium~\citep{Kong2004}. Following~\citet{Kong2002b}, we estimated the physical parameters of the SNR through the Sedov solution, assuming an initial explosion energy of $3\times10^{50}$~ergs \citep{Blair1981}, a radius of 9 pc, and an electron temperature of $2.37\pm1.24$ keV (average value from the two spectral fits). We derived the age of the SNR to be $2500^{+1100}_{-500}$~years and the number density of the ambient gas to be $0.10^{+0.14}_{-0.03}$~cm$^{-3}$. The derived age and number density are both relatively small compared to other X-ray observed SNRs. This could have explained why this SNR was not detected in the X-ray in previous missions as nondetected X-ray SNRs usually reside in regions with ambient densities less than 0.1~cm$^{-3}$~\citep{Magnier1997}.

\section{Acknowledgement}
We thank Dr. David Burrows and his collaborators for generously providing the source code to compute the Bayesian confidence intervals. We also thank the referee for very helpful comments and suggestions to improve the manuscript. This work is based on observations obtained with \xmm, an ESA science mission with instruments and contributions directly funded by ESA Member States and NASA. D.S.Y.~Mak acknowledges support from HKU under the grant of Postgraduate Studentship. C.S.J.~Pun acknowledges support of a RGC grant from the government of the Hong Kong SAR. A.K.H.~Kong acknowledges support from the National Science Council, Taiwan, through a grant NSC96-2112-M-007-037-MY3.


\begin{thebibliography}{64}
\expandafter\ifx\csname natexlab\endcsname\relax\def\natexlab#1{#1}\fi

\bibitem[{{Abolmasov} {et~al.}(2007){Abolmasov}, {Fabrika}, {Sholukhova}, \&
  {Afanasiev}}]{Abolmasov2007}
{Abolmasov}, P., {Fabrika}, S., {Sholukhova}, O., \& {Afanasiev}, V. 2007,
  Astrophysical Bulletin, 62, 36

\bibitem[{{Argiroffi} {et~al.}(2006){Argiroffi}, {Favata}, {Flaccomio},
  {Maggio}, {Micela}, {Peres}, \& {Sciortino}}]{Argiroffi2006}
{Argiroffi}, C., {Favata}, F., {Flaccomio}, E., {Maggio}, A., {Micela}, G.,
  {Peres}, G., \& {Sciortino}, S. 2006, \aap, 459, 199

\bibitem[{{Baker} {et~al.}(1977){Baker}, {Haslam}, {Wielebinski}, \&
  {Jones}}]{Baker1977}
{Baker}, J.~R., {Haslam}, C.~G.~T., {Wielebinski}, R., \& {Jones}, B.~B. 1977,
  \aap, 59, 261

\bibitem[{{Bauer} {et~al.}(2003){Bauer}, {Brandt}, \& {Lehmer}}]{Bauer2003}
{Bauer}, F.~E., {Brandt}, W.~N., \& {Lehmer}, B. 2003, \aj, 126, 2797

\bibitem[{{Belloni} {et~al.}(2000){Belloni}, {Klein-Wolt}, {M{\'e}ndez}, {van
  der Klis}, \& {van Paradijs}}]{Belloni2000}
{Belloni}, T., {Klein-Wolt}, M., {M{\'e}ndez}, M., {van der Klis}, M., \& {van
  Paradijs}, J. 2000, \aap, 355, 271

\bibitem[{{Blair} {et~al.}(1981){Blair}, {Kirshner}, \&
  {Chevalier}}]{Blair1981}
{Blair}, W.~P., {Kirshner}, R.~P., \& {Chevalier}, R.~A. 1981, \apj, 247, 879

\bibitem[{{B{\"o}ker} {et~al.}(1999){B{\"o}ker}, {van der Marel}, \&
  {Vacca}}]{Boker1999}
{B{\"o}ker}, T., {van der Marel}, R.~P., \& {Vacca}, W.~D. 1999, \aj, 118, 831

\bibitem[{{Brandt} {et~al.}(2001){Brandt}, {Alexander}, {Hornschemeier},
  {Garmire}, {Schneider}, {Barger}, {Bauer}, {Broos}, {Cowie}, {Townsley},
  {Burrows}, {Chartas}, {Feigelson}, {Griffiths}, {Nousek}, \&
  {Sargent}}]{Brandt2001}
{Brandt}, W.~N. {et~al.} 2001, \aj, 122, 2810

\bibitem[{{Bregman} {et~al.}(1993){Bregman}, {Cox}, \&
  {Tomisaka}}]{Bregman1993}
{Bregman}, J.~N., {Cox}, C.~V., \& {Tomisaka}, K. 1993, \apjl, 415, L79

\bibitem[{{Buta} \& {McCall}(1999)}]{Buta1999}
{Buta}, R.~J., \& {McCall}, M.~L. 1999, \apjs, 124, 33

\bibitem[{{Cutri} {et~al.}(2003){Cutri}, {Skrutskie}, {van Dyk}, {Beichman},
  {Carpenter}, {Chester}, {Cambresy}, {Evans}, {Fowler}, {Gizis}, {Howard},
  {Huchra}, {Jarrett}, {Kopan}, {Kirkpatrick}, {Light}, {Marsh}, {McCallon},
  {Schneider}, {Stiening}, {Sykes}, {Weinberg}, {Wheaton}, {Wheelock}, \&
  {Zacarias}}]{Cutri2003}
{Cutri}, R.~M. {et~al.} 2003, {2MASS All Sky Catalog of point sources.} (The
  IRSA 2MASS All-Sky Point Source Catalog, NASA/IPAC Infrared Science
  Archive.~http://irsa.ipac.caltech.edu/applications/Gator/)

\bibitem[{{de Vaucouleurs} {et~al.}(1992){de Vaucouleurs}, {de Vaucouleurs},
  {Corwin}, {Buta}, {Paturel}, \& {Fouque}}]{Vaucouleurs1992}
{de Vaucouleurs}, G., {de Vaucouleurs}, A., {Corwin}, Jr., H.~G., {Buta},
  R.~J., {Paturel}, G., \& {Fouque}, P. 1992, VizieR Online Data Catalog, 7137,
  0

\bibitem[{{Di Stefano} \& {Kong}(2003{\natexlab{a}})}]{Stefano2003a}
{Di Stefano}, R., \& {Kong}, A.~K.~H. 2003{\natexlab{a}}, \apj, 592, 884

\bibitem[{{Di Stefano} \& {Kong}(2003{\natexlab{b}})}]{Stefano2003b}
---. 2003{\natexlab{b}}, ArXiv Astrophysics e-prints

\bibitem[Di Stefano 
\& Kong (2004)]{StefanoKong2004} Di Stefano, R., \& Kong, A.~K.~H.\ 2004, \apj, 609, 710

\bibitem[{{Di Stefano} {et~al.}(2004){Di Stefano}, {Kong}, {Greiner},
  {Primini}, {Garcia}, {Barmby}, {Massey}, {Hodge}, {Williams}, {Murray},
  {Curry}, \& {Russo}}]{Stefano2004}
{Di Stefano}, R. {et~al.} 2004, \apj, 610, 247

\bibitem[{{Dodorico} {et~al.}(1980){Dodorico}, {Dopita}, \&
  {Benvenuti}}]{Dodorico1980}
{Dodorico}, S., {Dopita}, M.~A., \& {Benvenuti}, P. 1980, \aaps, 40, 67

\bibitem[{{Fabbiano}(2006)}]{Fabbiano2006}
{Fabbiano}, G. 2006, \araa, 44, 323

\bibitem[{{Fabbiano} \& {Trinchieri}(1987)}]{Fabbiano1987}
{Fabbiano}, G., \& {Trinchieri}, G. 1987, \apj, 315, 46

\bibitem[{{Favata} {et~al.}(1997){Favata}, {Vink}, {Parmar}, {Kaastra}, \&
  {Mineo}}]{Favata1997}
{Favata}, F., {Vink}, J., {Parmar}, A.~N., {Kaastra}, J.~S., \& {Mineo}, T.
  1997, \aap, 324, L45

\bibitem[{{Feng} \& {Kaaret}(2008)}]{Feng2008}
{Feng}, H., \& {Kaaret}, P. 2008, \apj, 675, 1067

\bibitem[{{Finlator} {et~al.}(2000){Finlator}, {Ivezi{\'c}}, {Fan}, {Strauss},
  {Knapp}, {Lupton}, {Gunn}, {Rockosi}, {Anderson}, {Csabai}, {Hennessy},
  {Hindsley}, {McKay}, {Nichol}, {Schneider}, {Smith}, {York}, \& {the SDSS
  Collaboration}}]{Finlator2000}
{Finlator}, K. {et~al.} 2000, \aj, 120, 2615

\bibitem[{{Freeman} {et~al.}(2002){Freeman}, {Kashyap}, {Rosner}, \&
  {Lamb}}]{Freeman2002}
{Freeman}, P.~E., {Kashyap}, V., {Rosner}, R., \& {Lamb}, D.~Q. 2002, \apjs,
  138, 185

\bibitem[{{Fridriksson} {et~al.}(2008){Fridriksson}, {Homan}, {Lewin}, {Kong},
  \& {Pooley}}]{Fridriksson2008}
{Fridriksson}, J.~K., {Homan}, J., {Lewin}, W.~H.~G., {Kong}, A.~K.~H., \&
  {Pooley}, D. 2008, ArXiv e-prints, 804

\bibitem[{{Greiner} {et~al.}(2004){Greiner}, {Di Stefano}, {Kong}, \&
  {Primini}}]{Greiner2004}
{Greiner}, J., {Di Stefano}, R., {Kong}, A., \& {Primini}, F. 2004, \apj, 610,
  261

\bibitem[{{Grimm} {et~al.}(2007){Grimm}, {McDowell}, {Zezas}, {Kim}, \&
  {Fabbiano}}]{Grimm2007}
{Grimm}, H.-J., {McDowell}, J., {Zezas}, A., {Kim}, D.-W., \& {Fabbiano}, G.
  2007, \apjs, 173, 70

\bibitem[{{Hughes} {et~al.}(1998){Hughes}, {Hayashi}, \& {Koyama}}]{Hughes1998}
{Hughes}, J.~P., {Hayashi}, I., \& {Koyama}, K. 1998, \apj, 505, 732

\bibitem[{{Irwin} {et~al.}(2003){Irwin}, {Athey}, \& {Bregman}}]{Irwin2003}
{Irwin}, J.~A., {Athey}, A.~E., \& {Bregman}, J.~N. 2003, \apj, 587, 356

\bibitem[{{Jim{\'e}nez-Bail{\'o}n} {et~al.}(2005){Jim{\'e}nez-Bail{\'o}n},
  {Santos-Lle{\'o}}, {Dahlem}, {Ehle}, {Mas-Hesse}, {Guainazzi}, {Heckman}, \&
  {Weaver}}]{Jimenez2005}
{Jim{\'e}nez-Bail{\'o}n}, E., {Santos-Lle{\'o}}, M., {Dahlem}, M., {Ehle}, M.,
  {Mas-Hesse}, J.~M., {Guainazzi}, M., {Heckman}, T.~M., \& {Weaver}, K.~A.
  2005, \aap, 442, 861

\bibitem[{{Kilgard} {et~al.}(2005){Kilgard}, {Cowan}, {Garcia}, {Kaaret},
  {Krauss}, {McDowell}, {Prestwich}, {Primini}, {Stockdale}, {Trinchieri},
  {Ward}, \& {Zezas}}]{Kilgard2005}
{Kilgard}, R.~E. {et~al.} 2005, \apjs, 159, 214

\bibitem[{{King} {et~al.}(2001){King}, {Davies}, {Ward}, {Fabbiano}, \&
  {Elvis}}]{King2001}
{King}, A.~R., {Davies}, M.~B., {Ward}, M.~J., {Fabbiano}, G., \& {Elvis}, M.
  2001, \apjl, 552, L109

\bibitem[{{Kong}(2003)}]{Kong2003}
{Kong}, A.~K.~H. 2003, \mnras, 346, 265

\bibitem[Kong et al.(2003)]{KongSNR2003} Kong, A.~K.~H., 
Sjouwerman, L.~O., Williams, B.~F., Garcia, M.~R., 
\& Dickel, J.~R.\ 2003, \apjl, 590, L21 

\bibitem[Kong 
\& Di Stefano(2003)]{KongStefano2003} Kong, A.~K.~H., \& Di Stefano, R.\ 2003, \apjl, 590, L13

\bibitem[{{Kong} {et~al.}(2004{\natexlab{a}}){Kong}, {Di Stefano}, \&
  {Yuan}}]{Kong2004b}
{Kong}, A.~K.~H., {Di Stefano}, R., \& {Yuan}, F. 2004{\natexlab{a}}, \apjl,
  617, L49
  
\bibitem[Kong 
\& Di Stefano (2005)]{Kong2005} Kong, A.~K.~H., \& Di Stefano, R.\ 2005, \apjl, 632, L107 


\bibitem[{{Kong} {et~al.}(2002{\natexlab{a}}){Kong}, {Garcia}, {Primini}, \&
  {Murray}}]{Kong2002b}
{Kong}, A.~K.~H., {Garcia}, M.~R., {Primini}, F.~A., \& {Murray}, S.~S.
  2002{\natexlab{a}}, \apjl, 580, L125

\bibitem[{{Kong} {et~al.}(2002{\natexlab{b}}){Kong}, {Garcia}, {Primini},
  {Murray}, {Di Stefano}, \& {McClintock}}]{Kong2002}
{Kong}, A.~K.~H., {Garcia}, M.~R., {Primini}, F.~A., {Murray}, S.~S., {Di
  Stefano}, R., \& {McClintock}, J.~E. 2002{\natexlab{b}}, \apj, 577, 738

\bibitem[{{Kong} {et~al.}(2004{\natexlab{b}}){Kong}, {Sjouwerman}, \&
  {Williams}}]{Kong2004}
{Kong}, A.~K.~H., {Sjouwerman}, L.~O., \& {Williams}, B.~F. 2004{\natexlab{b}},
  \aj, 128, 2783

\bibitem[{{K{\"o}rding} {et~al.}(2002){K{\"o}rding}, {Falcke}, \&
  {Markoff}}]{Kording2002}
{K{\"o}rding}, E., {Falcke}, H., \& {Markoff}, S. 2002, \aap, 382, L13

\bibitem[{{Kraft} {et~al.}(1991){Kraft}, {Burrows}, \& {Nousek}}]{Kraft1991}
{Kraft}, R.~P., {Burrows}, D.~N., \& {Nousek}, J.~A. 1991, \apj, 374, 344

\bibitem[{{Kubota} {et~al.}(2002){Kubota}, {Done}, \& {Makishima}}]{Kubota2002}
{Kubota}, A., {Done}, C., \& {Makishima}, K. 2002, \mnras, 337, L11

\bibitem[{{Kubota} {et~al.}(2001){Kubota}, {Mizuno}, {Makishima}, {Fukazawa},
  {Kotoku}, {Ohnishi}, \& {Tashiro}}]{Kubota2001}
{Kubota}, A., {Mizuno}, T., {Makishima}, K., {Fukazawa}, Y., {Kotoku}, J.,
  {Ohnishi}, T., \& {Tashiro}, M. 2001, \apjl, 547, L119

\bibitem[{{Maccacaro} {et~al.}(1988){Maccacaro}, {Gioia}, {Wolter}, {Zamorani},
  \& {Stocke}}]{Maccacaro1988}
{Maccacaro}, T., {Gioia}, I.~M., {Wolter}, A., {Zamorani}, G., \& {Stocke},
  J.~T. 1988, \apj, 326, 680

\bibitem[{{Magnier} {et~al.}(1997){Magnier}, {Primini}, {Prins}, {van
  Paradijs}, \& {Lewin}}]{Magnier1997}
{Magnier}, E.~A., {Primini}, F.~A., {Prins}, S., {van Paradijs}, J., \&
  {Lewin}, W.~H.~G. 1997, \apj, 490, 649

\bibitem[{{Mak} {et~al.}(2008){Mak}, {Pun}, \& {Kong}}]{Mak2007}
{Mak}, D.~S.~Y., {Pun}, C.~S.~J., \& {Kong}, A.~K.~H. 2008, \apj, 686, 995

\bibitem[{{Markoff} {et~al.}(2003){Markoff}, {Nowak}, {Corbel}, {Fender}, \&
  {Falcke}}]{Markoff2003}
{Markoff}, S., {Nowak}, M., {Corbel}, S., {Fender}, R., \& {Falcke}, H. 2003,
  \aap, 397, 645

\bibitem[{{Meier} \& {Turner}(2005)}]{Meier2005}
{Meier}, D.~S., \& {Turner}, J.~L. 2005, \apj, 618, 259

\bibitem[{{Misanovic} {et~al.}(2006){Misanovic}, {Pietsch}, {Haberl}, {Ehle},
  {Hatzidimitriou}, \& {Trinchieri}}]{Misanovic2006}
{Misanovic}, Z., {Pietsch}, W., {Haberl}, F., {Ehle}, M., {Hatzidimitriou}, D.,
  \& {Trinchieri}, G. 2006, \aap, 448, 1247

\bibitem[{{Mitsuda} {et~al.}(1984){Mitsuda}, {Inoue}, {Koyama}, {Makishima},
  {Matsuoka}, {Ogawara}, {Suzuki}, {Tanaka}, {Shibazaki}, \&
  {Hirano}}]{Mitsuda1984}
{Mitsuda}, K. {et~al.} 1984, \pasj, 36, 741

\bibitem[{{Monet} {et~al.}(2003){Monet}, {Levine}, {Canzian}, {Ables}, {Bird},
  {Dahn}, {Guetter}, {Harris}, {Henden}, {Leggett}, {Levison}, {Luginbuhl},
  {Martini}, {Monet}, {Munn}, {Pier}, {Rhodes}, {Riepe}, {Sell}, {Stone},
  {Vrba}, {Walker}, {Westerhout}, {Brucato}, {Reid}, {Schoening}, {Hartley},
  {Read}, \& {Tritton}}]{Monet2003}
{Monet}, D.~G. {et~al.} 2003, \aj, 125, 984

\bibitem[{{Newton}(1980)}]{Newton1980}
{Newton}, K. 1980, \mnras, 191, 169

\bibitem[{{Okada} {et~al.}(1998){Okada}, {Dotani}, {Makishima}, {Mitsuda}, \&
  {Mihara}}]{Okada1998}
{Okada}, K., {Dotani}, T., {Makishima}, K., {Mitsuda}, K., \& {Mihara}, T.
  1998, \pasj, 50, 25

\bibitem[{{Osborne} {et~al.}(2001){Osborne}, {Borozdin}, {Trudolyubov},
  {Priedhorsky}, {Soria}, {Shirey}, {Hayter}, {La Palombara}, {Mason},
  {Molendi}, {Paerels}, {Pietsch}, {Read}, {Tiengo}, {Watson}, \&
  {West}}]{Osborne2001}
{Osborne}, J.~P. {et~al.} 2001, \aap, 378, 800

\bibitem[{{Pakull} \& {Mirioni}(2002)}]{Pakull2002}
{Pakull}, M.~W., \& {Mirioni}, L. 2002, ArXiv Astrophysics e-prints

\bibitem[{{Primini} {et~al.}(1993){Primini}, {Forman}, \&
  {Jones}}]{Primini1993}
{Primini}, F.~A., {Forman}, W., \& {Jones}, C. 1993, \apj, 410, 615

\bibitem[{{Remillard} \& {McClintock}(2006)}]{Remillard2006}
{Remillard}, R.~A., \& {McClintock}, J.~E. 2006, \araa, 44, 49

\bibitem[{{Roberts} {et~al.}(2003){Roberts}, {Goad}, {Ward}, \&
  {Warwick}}]{Roberts2003}
{Roberts}, T.~P., {Goad}, M.~R., {Ward}, M.~J., \& {Warwick}, R.~S. 2003,
  \mnras, 342, 709

\bibitem[{{Schinnerer} {et~al.}(2003){Schinnerer}, {B{\"o}ker}, \&
  {Meier}}]{Schinnerer2003}
{Schinnerer}, E., {B{\"o}ker}, T., \& {Meier}, D.~S. 2003, \apjl, 591, L115

\bibitem[{{Schlegel} {et~al.}(2000){Schlegel}, {Blair}, \&
  {Fesen}}]{Schlegel2000}
{Schlegel}, E.~M., {Blair}, W.~P., \& {Fesen}, R.~A. 2000, \aj, 120, 791

\bibitem[{{Schulz} {et~al.}(2001){Schulz}, {G{\"u}sten}, {K{\"o}ster}, \&
  {Krause}}]{Schulz2001}
{Schulz}, A., {G{\"u}sten}, R., {K{\"o}ster}, B., \& {Krause}, D. 2001, \aap,
  371, 25

\bibitem[{{Sivakoff} {et~al.}(2008){Sivakoff}, {Jord'an}, {Juett}, {Sarazin},
  \& {Irwin}}]{Sivakoff2008}
{Sivakoff}, G.~R., {Jord'an}, A., {Juett}, A.~M., {Sarazin}, C.~L., \& {Irwin},
  J.~A. 2008, ArXiv e-prints, 806

\bibitem[{{Snowden}(2002)}]{Snowden2002}
{Snowden}, S.~L. 2002, ArXiv Astrophysics e-prints/0203311

\bibitem[{{Stark} {et~al.}(1992){Stark}, {Gammie}, {Wilson}, {Bally}, {Linke},
  {Heiles}, \& {Hurwitz}}]{Stark1992}
{Stark}, A.~A., {Gammie}, C.~F., {Wilson}, R.~W., {Bally}, J., {Linke}, R.~A.,
  {Heiles}, C., \& {Hurwitz}, M. 1992, \apjs, 79, 77

\bibitem[{{Sugiho} {et~al.}(2001){Sugiho}, {Kotoku}, {Makishima}, {Kubota},
  {Mizuno}, {Fukazawa}, \& {Tashiro}}]{Sugiho2001}
{Sugiho}, M., {Kotoku}, J., {Makishima}, K., {Kubota}, A., {Mizuno}, T.,
  {Fukazawa}, Y., \& {Tashiro}, M. 2001, \apjl, 561, L73

\bibitem[{{van den Heuvel} {et~al.}(1992){van den Heuvel}, {Bhattacharya},
  {Nomoto}, \& {Rappaport}}]{Heuvel1992}
{van den Heuvel}, E.~P.~J., {Bhattacharya}, D., {Nomoto}, K., \& {Rappaport},
  S.~A. 1992, \aap, 262, 97

\bibitem[{{Williams} {et~al.}(1999){Williams}, {Chu}, {Dickel}, {Petre},
  {Smith}, \& {Tavarez}}]{Williams1999}
{Williams}, R.~M., {Chu}, Y.-H., {Dickel}, J.~R., {Petre}, R., {Smith}, R.~C.,
  \& {Tavarez}, M. 1999, \apjs, 123, 467
  
\bibitem[Williams et al.(2005)]{Williams2005} Williams, B.~F., 
Barnard, R., Garcia, M.~R., Kolb, U., Osborne, J.~P., 
\& Kong, A.~K.~H.\ 2005, \apj, 634, 365 

\bibitem[Williams et al.(2007)]{Williams2007} Williams, B.~F., 
Barnard, R., Garcia, M.~R., Kolb, U., Osborne, J.~P., 
\& Kong, A.~K.~H.\ 2007, \apj, 666, 626 

\bibitem[{{Zezas} {et~al.}(2006){Zezas}, {Fabbiano}, {Baldi}, {Schweizer},
  {King}, {Ponman}, \& {Rots}}]{Zezas2006}
{Zezas}, A., {Fabbiano}, G., {Baldi}, A., {Schweizer}, F., {King}, A.~R.,
  {Ponman}, T.~J., \& {Rots}, A.~H. 2006, \apjs, 166, 211

\end{thebibliography}

\begin{landscape}
\begin{deluxetable}{ccccccc}
\tablecolumns{7}
\tablewidth{0pc}
\tablecaption{Archival \xmm\ Observations of IC342}
\tablehead{
\colhead{} & \colhead{} & \multicolumn{2}{c}{Aim point} & \multicolumn{3}{c}{$T_{exp}$ (ksec)}  \\
\colhead{Obs ID} & \colhead{Date} & \colhead{RA (J2000.0)} & \colhead{DEC (J2000.0)}  & \colhead{EPIC PN} & \colhead{EPIC MOS1}  & \colhead{EPIC MOS2} } 
\startdata
0093640901&2001-02-11&03:46:49.40&+68:05:38.9&5.6  &9.5  &9.6\\
0206890101&2004-02-20&03:46:15.92&+68:08:43.6&13.0&20.1&20.7\\
0206890201&2004-08-17&03:45:56.41&+68:07:23.7&19.7&23.6&23.5\\
0206890401&2005-02-10&03:45:55.66&+68:07:31.8&6.1  &6.1  &5.9\\
\enddata
\label{t:obs}
\end{deluxetable}
\end{landscape}

\begin{landscape}
\begin{deluxetable}{cccccccccccc}
\tablecolumns{12}
\tablewidth{0pc}
\tabletypesize{\scriptsize}
\tablecaption{\xmm\ X-ray source catalog of IC342}
\tablehead{
\colhead{ID} & \colhead{Source Name} & \colhead{RA} & \colhead{DEC} & \colhead{Positional error}  & \multicolumn{3}{c}{Net counts} & \colhead{HR1}  & \colhead{HR2} & \colhead{$L_{\rm X}$ (0.3--12 keV)}  & \colhead{Note} \\
\colhead{} & \colhead{XMMU J} & \colhead{(J2000.0)} & \colhead{(J2000.0)} & \colhead{(arcsec)}  & \colhead{soft (0.3--1 keV)} & \colhead{medium (1.0--2.0 keV)} & \colhead{hard (2.0--12.0 keV)}  & \colhead{} & \colhead{}& \colhead{($\times10^{38}\ \lum$)}  & \colhead{} 
}
\startdata
$1^{\ast}$ 	&	034333.4+680224	&	03:43:33.47		&	68:02:24.8 	&	1.34	&$	88.9	\pm	10.0	$ & $	181.2	\pm	14.0	$ & $	271.3	\pm	19.5	$ & $	0.34	\pm	0.07	$ & $	0.51	\pm	0.07	$ & $	0.12	\pm	0.02	$ &		\\
$2^{\ast}$ 	&	034354.8+675657	&	03:43:54.88		&	67:56:57.4	&	1.41	&$	733.1	\pm	31.9	$ & $	292.4	\pm	19.8	$ & $	83.3	\pm	29.5	$ & $	-0.43	\pm	0.04	$ & $	-0.80	\pm	0.07	$ & $	0.67	\pm	0.06	$ &						\\
3	&	034402.1+680935	&	03:44:02.13		&	68:09:35.8	&	0.73	&$	124.4	\pm	13.3	$ & $	391.7	\pm	21.8	$ & $	434.5	\pm	27.7	$ & $	0.52	\pm	0.06	$ & $	0.55	\pm	0.06	$ & $	0.55	\pm	0.04	$ &						\\
4	&	034426.8+681739	&	03:44:26.82		&	68:17:39.0 	&	0.85	&$	195.0	\pm	15.2	$ & $	74.9	\pm	9.8	$ & $	<8.9			$ & $	-0.45	\pm	0.07	$ & $	-0.75	\pm	0.11	$ & $	0.77	\pm	0.05	$ &	f	,	t			\\
5	&	034441.1+680827	&	03:44:41.10		&	68:08:27.7 	&	0.83	&$	<21.8			$ & $	114.9	\pm	13.0	$ & $	166.0	\pm	18.8	$ & $	0.95	\pm	0.17	$ & $	0.97	\pm	0.16	$ & $	0.10	\pm	0.01	$ &						\\
6	&	034447.4+680840	&	03:44:47.40		&	68:08:40.8 	&	0.63	&$	85.2	\pm	12.1	$ & $	426.6	\pm	22.3	$ & $	503.2	\pm	27.2	$ & $	0.67	\pm	0.06	$ & $	0.71	\pm	0.06	$ & $	0.50	\pm	0.05	$ &	f	,				\\
7	&	034449.0+680736	&	03:44:49.06		&	68:07:36.3 	&	1.34	&$	74.4	\pm	9.2	$ & $	<17.2			$ & $	<17.2			$ & $	-0.72	\pm	0.15	$ & $	-0.82	\pm	0.17	$ & $	0.31	\pm	0.04	$ &	f	,	t			\\
8	&	034449.6+680216	&	03:44:49.65		&	68:02:16.3 	&	0.80	&$	23.7	\pm	6.8	$ & $	80.6	\pm	10.6	$ & $	105.0	\pm	14.3	$ & $	0.55	\pm	0.14	$ & $	0.63	\pm	0.15	$ & $	0.16	\pm	0.03	$ &						\\
9	&	034504.5+675910	&	03:45:04.56		&	67:59:10.7 	&	1.49	&$	<26.5			$ & $	18.8	\pm	5.7	$ & $	33.4	\pm	8.8	$ & $	0.38	\pm	0.27	$ & $	0.60	\pm	0.27	$ & $	0.09	\pm	0.03	$ &	f	,			s	\\
10	&	034507.5+680111	&	03:45:07.51		&	68:01:11.7 	&	0.60	&$	102.0	\pm	10.7	$ & $	169.5	\pm	13.8	$ & $	115.3	\pm	12.7	$ & $	0.25	\pm	0.07	$ & $	0.06	\pm	0.08	$ & $	0.74	\pm	0.07	$ &	f	,	t			\\
11	&	034509.3+680855	&	03:45:09.32		&	68:08:55.7 	&	0.80	&$	32.7	\pm	8.6	$ & $	84.5	\pm	11.3	$ & $	80.0	\pm	16.0	$ & $	0.44	\pm	0.13	$ & $	0.42	\pm	0.17	$ & $	0.09	\pm	0.02	$ &	f	,				\\
12	&	034510.1+680231	&	03:45:10.12		&	68:02:31.0 	&	0.44	&$	253.5	\pm	17.1	$ & $	956.5	\pm	31.8	$ & $	875.8	\pm	32.4	$ & $	0.58	\pm	0.03	$ & $	0.55	\pm	0.04	$ & $	0.87	\pm	0.05	$ &	f	,	t			\\
13	&	034515.3+681648	&	03:45:15.37		&	68:16:48.5 	&	0.95	&$	<11.5			$ & $	23.7	\pm	8.2	$ & $	93.2	\pm	13.9	$ & $	0.44	\pm	0.32	$ & $	0.82	\pm	0.19	$ & $	0.21	\pm	0.03	$ &	f	,	t			\\
14	&	034531.3+681912	&	03:45:31.32		&	68:19:12.0 	&	0.68	&$	123.3	\pm	13.9	$ & $	520.0	\pm	24.8	$ & $	584.1	\pm	31.2	$ & $	0.62	\pm	0.05	$ & $	0.65	\pm	0.06	$ & $	0.72	\pm	0.05	$ &						\\
15	&	034533.9+675912	&	03:45:33.93		&	67:59:12.9 	&	0.51	&$	63.9	\pm	8.9	$ & $	304.7	\pm	18.6	$ & $	489.0	\pm	24.5	$ & $	0.65	\pm	0.07	$ & $	0.77	\pm	0.06	$ & $	0.36	\pm	0.03	$ &						\\
16	&	034536.9+680733	&	03:45:36.93		&	68:07:33.4 	&	0.90	&$	15.3	\pm	6.1	$ & $	53.0	\pm	9.1	$ & $	37.0	\pm	10.7	$ & $	0.55	\pm	0.18	$ & $	0.42	\pm	0.25	$ & $	0.07	\pm	0.02	$ &					s	\\
17	&	034539.9+680309	&	03:45:39.91		&	68:03:09.4 	&	0.29	&$	330.2	\pm	19.4	$ & $	1230.0	\pm	36.2	$ & $	1789.6	\pm	44.6	$ & $	0.58	\pm	0.03	$ & $	0.69	\pm	0.03	$ & $	1.56	\pm	0.07	$ &	f					\\
18	&	034547.1+680547	&	03:45:47.12		&	68:05:47.5 	&	0.61	&$	<34.3			$ & $	51.2	\pm	10.4	$ & $	71.6	\pm	13.6	$ & $	0.74	\pm	0.25	$ & $	0.81	\pm	0.24	$ & $	0.10	\pm	0.01	$ &	f	,	t			\\
19	&	034555.6+680455	&	03:45:55.62		&	68:04:55.9 	&	0.09	&$	3846.8	\pm	65.4	$ & $	26206.4	\pm	163.9	$ & $	32480.2	\pm	183.9	$ & $	0.74	\pm	0.01	$ & $	0.79	\pm	0.01	$ & $	21.00	\pm	0.18	$ &	f	,			s	\\
20	&	034556.8+675925	&	03:45:56.86		&	67:59:25.8 	&	0.76	&$	45.6	\pm	7.9	$ & $	91.3	\pm	11.1	$ & $	156.5	\pm	15.5	$ & $	0.33	\pm	0.11	$ & $	0.55	\pm	0.10	$ & $	0.19	\pm	0.03	$ &	f	,	t	,	s	\\
21	&	034559.4+680537	&	03:45:59.44		&	68:05:37.6 	&	0.29	&$	105.9	\pm	12.3	$ & $	555.2	\pm	26.7	$ & $	799.6	\pm	31.9	$ & $	0.68	\pm	0.05	$ & $	0.77	\pm	0.05	$ & $	0.83	\pm	0.04	$ &						\\
22	&	034602.0+675513	&	03:46:02.09		&	67:55:13.7 	&	1.11	&$	<30.7			$ & $	20.9	\pm	6.1	$ & $	27.5	\pm	8.9	$ & $	0.35	\pm	0.26	$ & $	0.47	\pm	0.29	$ & $	0.14	\pm	0.02	$ &						\\
23	&	034606.1+681029	&	03:46:06.19		&	68:10:29.3 	&	0.43	&$	41.0	\pm	8.6	$ & $	228.6	\pm	16.8	$ & $	286.4	\pm	22.1	$ & $	0.70	\pm	0.09	$ & $	0.75	\pm	0.09	$ & $	0.30	\pm	0.02	$ &	s					\\
24	&	034606.5+680705	&	03:46:06.54		&	68:07:05.3 	&	0.33	&$	976.2	\pm	32.1	$ & $	358.8	\pm	19.8	$ & $	48.3	\pm	10.5	$ & $	-0.46	\pm	0.03	$ & $	-0.91	\pm	0.04	$ & $	3.24	\pm	0.10	$ &	f	,	s			\\
25	&	034615.6+681112	&	03:46:15.64		&	68:11:12.2 	&	0.12	&$	513.4	\pm	28.7	$ & $	9198.3	\pm	98.2	$ & $	41562.4	\pm	207.3	$ & $	0.89	\pm	0.01	$ & $	0.98	\pm	0.01	$ & $	12.00	\pm	0.11	$ &	f	,	s			\\
26	&	034619.5+680554	&	03:46:19.59		&	68:05:54.7 	&	0.83	&$	26.0	\pm	6.0	$ & $	25.4	\pm	6.8	$ & $	12.9	\pm	7.8	$ & $	-0.01	\pm	0.18	$ & $	-0.34	\pm	0.27	$ & $	0.10	\pm	0.02	$ &	f					\\
27	&	034622.1+680506	&	03:46:22.12		&	68:05:06.9 	&	0.55	&$	42.2	\pm	7.9	$ & $	262.5	\pm	17.0	$ & $	154.6	\pm	14.3	$ & $	0.72	\pm	0.08	$ & $	0.57	\pm	0.10	$ & $	0.18	\pm	0.02	$ &					s	\\
28	&	034625.9+680421	&	03:46:25.91		&	68:04:21.8 	&	0.75	&$	30.1	\pm	6.1	$ & $	24.1	\pm	5.7	$ & $	<22.0			$ & $	-0.11	\pm	0.16	$ & $	-0.43	\pm	0.21	$ & $	0.10	\pm	0.03	$ &	f					\\
29	&	034626.7+680455	&	03:46:26.70		&	68:04:55.4 	&	0.75	&$	52.6	\pm	7.9	$ & $	42.1	\pm	7.5	$ & $	18.1	\pm	6.6	$ & $	-0.11	\pm	0.12	$ & $	-0.49	\pm	0.16	$ & $	0.18	\pm	0.03	$ &	f	,	t			\\
30	&	034627.5+680410	&	03:46:27.56		&	68:04:10.7 	&	0.82	&$	27.8	\pm	5.7	$ & $	37.2	\pm	6.5	$ & $	<7.2			$ & $	0.14	\pm	0.13	$ & $	-0.61	\pm	0.24	$ & $	0.19	\pm	0.03	$ &	f	,	t	,	s	\\
31	&	034633.4+675855	&	03:46:33.48		&	67:58:55.8 	&	0.87	&$	31.7	\pm	7.0	$ & $	98.9	\pm	10.9	$ & $	139.1	\pm	14.7	$ & $	0.52	\pm	0.11	$ & $	0.63	\pm	0.11	$ & $	0.11	\pm	0.02	$ &						\\
32	&	034638.9+675551	&	03:46:38.94		&	67:55:51.6 	&	0.79	&$	53.9	\pm	8.4	$ & $	158.7	\pm	13.4	$ & $	106.0	\pm	13.0	$ & $	0.49	\pm	0.08	$ & $	0.33	\pm	0.10	$ & $	0.38	\pm	0.01	$ &	f	,			s	\\
33	&	034639.8+680517	&	03:46:39.80		&	68:05:17.5 	&	0.59	&$	<9.6			$ & $	<7.9			$ & $	256.5	\pm	19.0	$ & $	0.00	\pm	0.00	$ & $	1.00	\pm	0.11	$ & $	0.23	\pm	0.03	$ &	f					\\
34	&	034643.6+680611	&	03:46:43.65		&	68:06:11.4 	&	0.33	&$	<12.9			$ & $	21.3	\pm	7.5	$ & $	49.1	\pm	10.3	$ & $	0.49	\pm	0.37	$ & $	0.74	\pm	0.26	$ & $	1.22	\pm	0.05	$ &					s	\\
35	&	034644.0+681104	&	03:46:44.07		&	68:11:04.6 	&	0.80	&$	189.3	\pm	22.1	$ & $	729.3	\pm	37.2	$ & $	812.1	\pm	38.9	$ & $	0.59	\pm	0.05	$ & $	0.62	\pm	0.05	$ & $	0.07	\pm	0.01	$ &	f					\\
36	&	034645.1+680946	&	03:46:45.14		&	68:09:46.1 	&	0.22	&$	2045.2	\pm	46.1	$ & $	1642.4	\pm	41.5	$ & $	307.2	\pm	18.9	$ & $	-0.11	\pm	0.02	$ & $	-0.74	\pm	0.03	$ & $	4.93	\pm	0.10	$ &	f	,	t	,	s	\\
37	&	034646.1+680524	&	03:46:46.19		&	68:05:24.7 	&	0.22	&$	62.2	\pm	8.6	$ & $	63.7	\pm	9.1	$ & $	<9.5			$ & $	0.01	\pm	0.10	$ & $	-0.96	\pm	0.22	$ & $	1.22	\pm	0.07	$ &	f	,	t			\\
38	&	034648.5+680547	&	03:46:48.52		&	68:05:47.4 	&	0.20	&$	3517.3	\pm	63.0	$ & $	3931.4	\pm	68.0	$ & $	1497.1	\pm	45.4	$ & $	0.06	\pm	0.01	$ & $	-0.40	\pm	0.02	$ & $	9.63	\pm	0.16	$ &					s	\\
39	&	034651.3+680028	&	03:46:51.32		&	68:00:28.6 	&	0.54	&$	36.4	\pm	8.3	$ & $	277.6	\pm	18.2	$ & $	477.2	\pm	24.8	$ & $	0.77	\pm	0.08	$ & $	0.86	\pm	0.07	$ & $	0.37	\pm	0.03	$ &	f	,				\\
40	&	034652.5+680535	&	03:46:52.59		&	68:05:35.4 	&	0.46	&$	64.1	\pm	12.3	$ & $	74.6	\pm	13.9	$ & $	49.3	\pm	13.4	$ & $	0.08	\pm	0.13	$ & $	-0.13	\pm	0.16	$ & $	1.07	\pm	0.07	$ &	f	,	t	,	s	\\
41	&	034652.6+680847	&	03:46:52.63		&	68:08:47.2 	&	0.64	&$	<21.2			$ & $	67.9	\pm	9.7	$ & $	130.0	\pm	13.8	$ & $	0.86	\pm	0.20	$ & $	0.92	\pm	0.15	$ & $	0.10	\pm	0.01	$ &	f	,	t			\\
42	&	034652.7+680504	&	03:46:52.74		&	68:05:04.9 	&	0.69	&$	43.9	\pm	9.9	$ & $	150.4	\pm	14.7	$ & $	213.4	\pm	17.4	$ & $	0.55	\pm	0.10	$ & $	0.66	\pm	0.09	$ & $	0.26	\pm	0.03	$ &	f	,	t			\\
43	&	034654.3+675901	&	03:46:54.36		&	67:59:01.2 	&	0.86	&$	38.7	\pm	7.2	$ & $	80.5	\pm	10.0	$ & $	<23.7			$ & $	0.35	\pm	0.11	$ & $	-0.56	\pm	0.23	$ & $	0.17	\pm	0.03	$ &	f	,	t	,	s	\\
44	&	034657.2+680619	&	03:46:57.27		&	68:06:19.4 	&	0.23	&$	495.6	\pm	23.8	$ & $	1807.4	\pm	43.9	$ & $	1909.0	\pm	45.4	$ & $	0.57	\pm	0.02	$ & $	0.59	\pm	0.02	$ & $	2.93	\pm	0.08	$ &	f					\\
45	&	034659.3+680316	&	03:46:59.31		&	68:03:16.0 	&	0.54	&$	300.1	\pm	18.4	$ & $	914.7	\pm	31.4	$ & $	565.7	\pm	26.0	$ & $	0.51	\pm	0.03	$ & $	0.31	\pm	0.04	$ & $	0.80	\pm	0.03	$ &	f	,	t	,	s	\\
46	&	034701.9+680237	&	03:47:01.93		&	68:02:37.6 	&	0.52	&$	83.7	\pm	11.9	$ & $	261.0	\pm	18.3	$ & $	359.2	\pm	21.9	$ & $	0.51	\pm	0.07	$ & $	0.62	\pm	0.07	$ & $	0.44	\pm	0.03	$ &						\\
47	&	034703.8+680905	&	03:47:03.83		&	68:09:05.0 	&	0.58	&$	68.5	\pm	10.0	$ & $	151.7	\pm	13.3	$ & $	239.9	\pm	17.4	$ & $	0.38	\pm	0.08	$ & $	0.56	\pm	0.07	$ & $	0.30	\pm	0.02	$ &						\\
48	&	034704.6+680517	&	03:47:04.69		&	68:05:17.4 	&	0.68	&$	141.6	\pm	13.3	$ & $	102.4	\pm	12.0	$ & $	<35.7			$ & $	-0.16	\pm	0.07	$ & $	-0.70	\pm	0.12	$ & $	0.33	\pm	0.03	$ &						\\
49	&	034708.1+681059	&	03:47:08.12		&	68:10:59.0 	&	0.64	&$	88.0	\pm	10.1	$ & $	299.2	\pm	18.2	$ & $	231.5	\pm	17.7	$ & $	0.55	\pm	0.06	$ & $	0.45	\pm	0.07	$ & $	0.69	\pm	0.06	$ &	f	,	t			\\
50	&	034718.6+681128	&	03:47:18.63		&	68:11:28.5 	&	0.51	&$	151.4	\pm	13.8	$ & $	455.4	\pm	22.6	$ & $	468.9	\pm	25.0	$ & $	0.50	\pm	0.05	$ & $	0.51	\pm	0.05	$ & $	0.79	\pm	0.05	$ &	f					\\
51	&	034722.3+681538	&	03:47:22.38		&	68:15:38.5 	&	0.84	&$	61.0	\pm	8.5	$ & $	46.6	\pm	7.9	$ & $	<8.6			$ & $	-0.13	\pm	0.11	$ & $	-0.64	\pm	0.18	$ & $	0.31	\pm	0.04	$ &	f	,	t	,	s	\\
52	&	034722.9+680859	&	03:47:22.96		&	68:08:59.5 	&	0.40	&$	317.2	\pm	19.0	$ & $	1033.8	\pm	33.1	$ & $	375.5	\pm	22.7	$ & $	0.53	\pm	0.03	$ & $	0.08	\pm	0.04	$ & $	0.95	\pm	0.05	$ &	f	,	t	,	s	\\
$53^{\ast}$	&	034726.5+680849	&	03:47:26.58		&	68:08:49.2 	&	0.94	&$	<0			$ & $	<0			$ & $	88.9	\pm	10.4	$ & $	0.07	\pm	0.30	$ & $	0.81	\pm	0.15	$ & $	0.12	\pm	0.02	$ &	f	,	t			\\
54	&	034748.1+681530	&	03:47:48.12		&	68:15:30.1 	&	0.91	&$	23.4	\pm	6.8	$ & $	49.2	\pm	8.8	$ & $	41.3	\pm	12.8	$ & $	0.35	\pm	0.16	$ & $	0.28	\pm	0.23	$ & $	0.11	\pm	0.03	$ &	f	,	t			\\
55	&	034803.5+681114	&	03:48:03.58		&	68:11:14.9 	&	1.20	&$	<6.4			$ & $	36.1	\pm	7.9	$ & $	57.0	\pm	11.7	$ & $	0.93	\pm	0.32	$ & $	0.95	\pm	0.29	$ & $	0.04	\pm	0.02	$ &					s	\\
56	&	034805.1+680137	&	03:48:05.10		&	68:01:37.9 	&	1.37	&$	27.6	\pm	6.1	$ & $	16.9	\pm	4.9	$ & $	<8.1			$ & $	-0.24	\pm	0.18	$ & $	-1.00	\pm	0.41	$ & $	0.15	\pm	0.04	$ &	f	,	t	,	s	\\
57	&	034807.1+680455	&	03:48:07.10		&	68:04:55.5 	&	0.34	&$	329.3	\pm	22.2	$ & $	1202.9	\pm	39.4	$ & $	1857.1	\pm	52.0	$ & $	0.57	\pm	0.03	$ & $	0.70	\pm	0.03	$ & $	1.57	\pm	0.07	$ &	f	,				\\
58	&	034817.9+680204	&	03:48:17.99		&	68:02:04.4 	&	0.99	&$	44.9	\pm	11.1	$ & $	191.9	\pm	19.3	$ & $	399.1	\pm	30.6	$ & $	0.62	\pm	0.11	$ & $	0.80	\pm	0.09	$ & $	0.36	\pm	0.02	$ &						\\
\tablebreak
59	&	034825.4+680817	&	03:48:25.42		&	68:08:17.1 	&	1.46	&$	19.4	\pm	5.5	$ & $	38.3	\pm	9.0	$ & $	85.6	\pm	14.3	$ & $	0.33	\pm	0.19	$ & $	0.63	\pm	0.17	$ & $	0.14	\pm	0.03	$ &	f	,	t			\\
60	&	034832.3+675644	&	03:48:32.37		&	67:56:44.3 	&	2.00	&$	<9.3			$ & $	17.1	\pm	6.6	$ & $	64.0	\pm	10.6	$ & $	1.00	\pm	0.63	$ & $	1.00	\pm	0.25	$ & $	0.23	\pm	0.02	$ &						\\
61	&	034914.5+675732	&	03:49:14.58		&	67:57:32.4 	&	1.23	&$	19.5	\pm	7.0	$ & $	49.8	\pm	11.7	$ & $	242.4	\pm	17.5	$ & $	0.44	\pm	0.21	$ & $	0.85	\pm	0.09	$ & $	0.19	\pm	0.05	$ &						\\
\enddata
\tablecomments{
Column (1) lists the running ID number of the X-ray sources.
Column (2) lists the X-ray source names given as XMMU JHHMMSS.S+DDMMSS.
Column (3) and (4) list the X-ray source positions that used the astrometric reference as discussed in Section~\ref{ss:det}. 
Column (5) lists the rms positional errors returned by the WAVDETECT task after aligning the four observations. The cross registration errors of aligning the observations and the astrometric registration errors with the USNO B1.0 catalog are not included.
Column (6)--(8) list the total background and exposure corrected net counts and its errors of X-ray sources for all 4 observations in the soft, medium and hard band respectively, as determined by DMEXTRACT. For sources that have S/N $< 2.5$ in a particular band, the 3$\sigma$ upper limit are used instead.
Column (9) and (10) list the average HR1 and HR2 values, defined as HR1~$=(M-S)/(M+S)$ and HR2~$=(H-S)/(H+S)$, and their corresponding errors averaged from data sets where the the source was present.
Column (11) lists the unabsorbed $0.3-12$~keV luminosities in unit of ($\times10^{38}\lum$) determined by EMLDETECT, by assuming an absorbed power-law spectrum with $\Gamma=2$ and $N_{\rm H}=8\times10^{21}\rm cm^{-2}$, averaged from data sets where the source was present. A distance of 1.8 Mpc to IC342 is assumed. 
Column (12): Symbols are defined as follows: t = X-ray transient; f = flux variable; s = spectral variable. \\
$^{\ast}$ These sources were only detected with EDETECT\_CHAIN software of SASDAS and not by the WAVDETECT task of CIAO. 
}
\label{t:cat}
\end{deluxetable}
\end{landscape}

\begin{deluxetable}{cclll}
\tablecolumns{5}
\tablewidth{0pc}
\tabletypesize{\scriptsize}
\tablecaption{Cross-correlation of our \xmm\ catalog with \chandra, \rosat, \einstein, and previous \xmm\ catalogs}
\tablehead{
\colhead{Source ID}& \colhead{previous \xmm\ $^{a}$ (offset $''$)} & \colhead{\chandra\ $^{b}$ (offset $''$)} & \colhead{\rosat\ $^{c}$ (offset $''$)} & \colhead{\einstein\ $^{d}$ (offset $''$)}  
} 
\startdata
6	&	X1	($	3.2''	$)&		-		&		-		&		-		\\
7	&	X3	($	5.6''	$)&		-		&		-		&		-		\\
8	&	X2	($	1.6''	$)&		-		&		-		&		-		\\
9	&	X4	($	4.2''	$)&		-		&		-		&		-		\\
12&	-			&	C1	($	5.7''	$)&		-		&		-		\\
15&	X5	($	3.2''	$)&		-		&		-		&		-		\\
17&	X6	($	1.8''	$)&	C2	($	3.5''	$)&	R1	($	1.0''	$)&		-		\\
19&	X7	($	2.3''	$)&	C3	($	0.1''	$)&	R3	($	1.6''	$)&	IC342 X-1	($	25.2''$)	\\
20&	X8	($	2.6''	$)&		-		&		-		&		-		\\
21&	X9	($	3.4''	$)&	C4	($	1.1''	$)&		-		&		-		\\
22&	X10	($	4.2''	$)&		-		&		-		&		-		\\
23&	X11	($	2.3''	$)&		-		&		-		&		-		\\
24&	X12	($	1.2''	$)&	C5	($	1.2''	$)&	R4	($	0.4''	$)&		-		\\
25&	X13	($	3.7''	$)&	C6	($	2.4''	$)&	R5	($	7.3''	$)&	IC342 X-2	($	30.6''$)	\\
27&	X14	($	3.1''	$)&	C7	($	0.9''	$)&		-		&		-		\\
28&	X15	($	6.8''	$)&		-		&		-		&		-		\\
29&	X16	($	2.4''	$)&		-		&		-		&		-		\\
30&	 -&	C8	($	3.2''	$)&		-		&		-		\\
31&	X17	($	3.6''	$)&		-		&		-		&		-		\\
32&	X18	($	3.7''	$)&		-		&		-		&		-		\\
35&	X19	($	2.0''	$)&	C9	($	0.7''	$)&	R6	($	1.6''	$)&		-		\\
36&	X20	($	3.5''	$)&	C11	($	2.0''	$)&	R7	($	5.2''	$)&		-		\\
38&	X21	($	1.7''	$)&	C12	($	1.1''	$)&	R8	($	3.1''	$)&	IC342 X-3	($	40.7''$)	\\
39&	X22	($	2.2''	$)&		-		&		-		&		-		\\
41&	X23	($	4.8''	$)&		-		&		-		&		-		\\
42&	X24	($	2.1''	$)&	C14	($	1.1''	$)&		-		&		-		\\
43&	X25	($	5.0''	$)&		-		&		-		&		-		\\
44&	X26	($	2.1''	$)&	C15	($	1.1''	$)&	R9	($	4.5''	$)&		-		\\
46&	X27	($	0.9''	$)&	C16	($	0.7''	$)&		-		&		-		\\
47&	X28	($	1.5''	$)&	C17	($	3.2''	$)&		-		&		-		\\
48&	X29	($	3.5''	$)&	C18	($	1.2''	$)&		-		&		-		\\
50&	X30	($	3.4''	$)&	C21	($	0.8''	$)&		-		&		-		\\
52&	X31	($	2.2''	$)&	C22	($	1.1''	$)&		-		&		-		\\
56&	X32	($	5.7''	$)&		-		&		-		&		-		\\
57&	X33	($	3.1''	$)&	C23	($	2.9''	$)&	R10	($	6.5''	$)&		-		\\
58&	X34	($	7.1''	$)&		-		&		-		&		-		\\
59&	X35	($	3.9''	$)&		-		&		-		&		-		\\
60&	X36	($	5.5''	$)&		-		&		-		&		-		\\
61&	X37	($	0.9''	$)&		-		&		-		&		-		\\

\enddata
\tablecomments{
$^{a}$ Table 1 of~\citet{Kong2003}.~\citet{Bauer2003} also gave the analysis of the same \xmm\ observation as~\citet{Kong2003}, but the catalogs are similar and~\citet{Kong2003} has two more sources. Searching radius = $6''$\\
$^{b}$ Table 1 of~\citet{Mak2007}. Prefix "C" denotes \chandra\ sources. Searching radius = $6''$\\
$^{c}$ Table 1 of~\citet{Bregman1993}. Prefix "R" denotes \rosat\ sources. Searching radius = $6''$\\
$^{d}$ Table 3 of~\citet*{Fabbiano1987}. Prefix "IC342-X" denotes the 3 brightest ULX in IC342 identified by \einstein. Searching radius = $40''$
}
\label{t:xso}
\end{deluxetable}

\begin{deluxetable}{cccc}
\tablecolumns{4}
\tablewidth{0pc}
\tabletypesize{\scriptsize}
\tablecaption{Energy Conversion Factors for each instrument and energy band}
\tablehead{
\colhead{} & \multicolumn{3}{c}{ECF ($10^{-11}$)} \\
\colhead{Detector} & \colhead{Soft (0.3--1 keV)} & \colhead{Medium (1--2 keV)} & \colhead{Hard (2--12 keV)}
} 
\startdata
MOS1&0.032 &0.597&0.266 \\
MOS2&0.032 &0.597&0.266 \\
PN  &0.0987&1.24 &0.516 \\
\enddata
\tablecomments{
Count rate to flux conversion factors for each EPIC instruments in the soft, medium, and hard band, assuming an absorbed power-law spectrum with a photon index of 2 and absorption column density of $8\times10^{21}\rm cm^{-2}$.
}
\label{t:ecf}
\end{deluxetable}

\begin{landscape}
\begin{deluxetable}{ccccl}
\tablecolumns{5}
\tablewidth{0pc}
\tabletypesize{\scriptsize}
\tablecaption{Classification Scheme of the X-ray sources}
\tablehead{
\colhead{Object Type} & \multicolumn{4}{c}{Selection criteria} \\
\colhead{} & \colhead{Hardness Ratio} & \colhead{Catalog}  & \colhead{X-ray Spectral properties} & \colhead{Others} 
}
\startdata

XRB&-&$^{a}$ &XRB: diskbb$^{1}$ with $kT\approx1-3$ keV&XRBs: variable\\
SSS                       &c.f. fig. 3 in D03$^{b}$&-&bb with $kT\le100$eV&-\\
SNR                       & $HR1_{\rm z} > 0.1$ and $HR2_{\rm z} < -0.4$ (Z06)$^{c}$& Do80$^{d}$&dominated by thermal spectrum $<2$ keV&- \\
Stars          & $HR1_{\rm z} < 0.3$ and $HR2_{\rm z} < -0.4$&USNOB1.0$^{e}$& RS$^{2}$ with $kT\approx1$ keV & $J-K<0.8$ and $J<12.5$, $\log(f_{\rm X}/f_{\rm opt})\leq1$ \\
X-ray transient&-&-&-&1.variability factor > 3 \\
& & & &2. at least detected in one observation \\
& & & &with $L_{\rm X}>10^{37}\lum$ and not detected in\\
& & & &at least one other observation\\
\enddata
\tablecomments{
$^{1}$ Disk Blackbody model; $^{2}$ Raymond-Smith model\\
$^{a}\xmm:~$\citealt{Kong2003}; \chandra:~\citealt*{Mak2007}; \rosat:~\citealt*{Bregman1993}; \einstein:~\citealt*{Fabbiano1987};\\ $^{b}$\citealt*{Stefano2003b}; \\
$^{c}$\citealt{Misanovic2006};\\ $^{d}$\citealt{Dodorico1980}; \\$^{e}$\citealt{Monet2003}\\
}
\label{t:sch}
\end{deluxetable}
\end{landscape}

\begin{deluxetable}{clccccc}
\tablecolumns{7}
\tablewidth{0pc}
\tabletypesize{\scriptsize}
\tablecaption{Summary of the identified SSS (source 36) and QSS (sources 52, 54, 56)}
\tablehead{
\colhead{Source ID} & \colhead{Observation} & \colhead{Soft (0.1-1.1 keV)} &\colhead{Medium (1.1-2 keV)} & \colhead{Hard (2-7 keV)} & \colhead{HR1} & \colhead{HR2} \\
} 
\startdata
36         & 2002 February $\ast$ &$170.89\pm13.35$&$6.81\pm2.87$&$3.05\pm2.57$&-0.92 &-0.96 \\
             & 2004 February &$627.06\pm25.44$&$132.65\pm12.13$&$18.82\pm5.77$&-0.65&-0.94\\
             &2004 August &$1738.74\pm42.56$&$1047.46\pm33.09$&$270.79\pm16.86$&-0.25&-0.73\\
             &2005 February$\ast$&$	18.65	\pm	4.76	$&$	1.48	\pm	2.13	$&$	0.00	\pm	1.99	$&	-0.85	&	-1.00	\\	
\hline \hline
52	& 2002 February & $66.3\pm8.6$&$125.6\pm11.4$&$51.1\pm7.7$&0.31&-0.13\\
          & 2004 February &$216.1\pm15.6$&$	311.1\pm18.5$&$	109.8\pm12.3$&0.18&	-0.33\\
          &2004 August &$	206.8	\pm	15.3	$&$	434.8	\pm	21.4	$&$	159.4	\pm	13.5	$&	0.36	&	-0.13	\\
          &2005 February $\ast$&$	22.1	\pm	5.9	$&$	14.2	\pm	4.5	$&$	0.0	\pm	4.5	$&	-0.22	&	-1.00	\\
          \hline
54	& 2002 February & $3.4\pm3.9$&$0.0\pm2.9$&$0.0\pm3.6$&-1.00&-1.00\\
          & 2004 February$\ast$ &$ 2.0\pm8.9$&$42.0\pm9.6$&$0.0\pm10.6$&0.91&-1.00 \\
          &2004 August &$	22.6	\pm	7.5	$&$	48.8	\pm	8.6	$&$	25.8	\pm	8.5	$&	0.37	&	0.07	\\
          &2005 February$\ast$&$	11.4	\pm	6.4	$&$	6.6	\pm	5.1	$&$	0.0	\pm	5.4	$&	-0.27	&	-1.00	\\
          \hline
56	& 2002 February$\ast$ & $26.8\pm6.3$&$14.9\pm4.6$&$0.0\pm3.4$&	-0.29&-1.00\\
          & 2004 February &$2.4\pm9.1$&$2.4\pm7.6$&$19.5\pm11.0$&0.00&	0.78 \\
          &2004 August &$	15.7	\pm	7.8	$&$	10.8	\pm	7.2	$&$	9.1	\pm	6.8	$&	-0.18	&	-0.27	\\
          &2005 February&$	3.3	\pm	5.4	$&$	0.7	\pm	5.1	$&$	1.2	\pm	7.4	$&	-0.64	&	-0.47	\\	
\enddata
\tablecomments{
$^{\ast}$ Observations in which the source satisfied the selection criteria of SSS or QSS. \\
}
\label{t:sss}
\end{deluxetable}

\begin{deluxetable}{cccccc}
\tablecolumns{6}
\tablewidth{0pc}
\tabletypesize{\scriptsize}
\tablecaption{Identification of SSS, QSS, SNR, and foreground star}
\tablehead{
\colhead{Source ID} & \colhead{Type$^{1}$} & \colhead{Identification$^{2}$} & \multicolumn{3}{c}{Radial Offset$^{3}$} \\
\colhead{} & \colhead{} & \colhead{} & \colhead{USNOB1.0} & \colhead{2MASS} & \colhead{D08$^{2}$}
} 
\startdata
2&Star&Cat3&$4.0''$&$3.9''$ &-\\ 
7&Star&Cat1&$4.0''$&$0.7''$&-\\ 
16&Star&Cat2&$3.1''$&$3.3''$&- \\ 
24&Star&Cat1&$1.5''$&$1.2''$&- \\ 
26&Star&Cat1&$3.1''$&$3.1''$&-\\ 
27&SNR&D80&-&-&$9.4''$\\ 
28&Star&Cat2&$4.3''$&$5.2''$&-\\ 
29&Star&Cat1&$1.7''$&$2.2''$&- \\ 
36&SSS&HR&-&-&-\\ 
37&Star&Cat2&$1.8''$&-&-\\ 
43&Star&Cat2&$2.5''$&$2.2''$&-\\ 
48&SNR&HR&$0.6''$&-&-\\ 
51&Star&Cat2&$1.5''$&$2.2''$&-\\ 
52$^\ast$&QSS-noh &HR&-&-&-\\ 
54$^\ast$&QSS-noh&HR&-&-&-\\ 
56&QSS-noh&HR&-&-&-\\ 
61&Star&Cat2&$2.6''$&$3.1''$&- \\ 
\enddata
\tablecomments{
$^{\ast}$ These sources are classified as SSS/QSS candidates only.\\
$^{1}$ Classification of SSS and QSS adopted the scheme by~\citet*{Stefano2003b}. QSS-noh: sources exhibited little or no emission above 1 keV; \\
$^{2}$ Foreground star classification (Cat1: USNOB1.0 + X-ray to optical flux ratio + 2MASS + hardness ratio; Cat2: USNOB1.0 + X-ray to optical flux ratio + hardness ratio; Cat3: USNOB1.0 + X-ray to optical flux ratio + 2Mass); D80:  \citealt{Dodorico1980}; HR: hardness ratio criteria\\
$^{3}$ Searching radius of catalogs: USNOB1.0 = $5''$; 2MASS = $5''$ D80 = $10''$
}
\label{t:ide}
\end{deluxetable}

\begin{deluxetable}{lccccccc}
\tablecolumns{8}
\tablewidth{0pc}
\tabletypesize{\scriptsize}
\tablecaption{Spectral fits to the brightest 11 X-ray sources in IC342}
\tablehead{
\colhead{Source ID} & \colhead{Obs year} & \colhead{XSPEC model} & \colhead{$N_{\rm H}~(10^{22}\rm cm^{-2})$} & \colhead{$\Gamma$ / $\log(n_{\rm e}t$)$^{a}$
} & \colhead{$kT^{b}$ (keV)} & \colhead{$\chi^{2}_{\upsilon}$/dof} & \colhead{$L_{\rm X}^{c}~(10^{38}\lum)$} \\
} 
\startdata
\cutinhead{Flux variables$^{\dagger}$}
17&2004 February&pow&$0.39^{+0.33}_{-0.26}$&$1.26^{+0.30}_{-0.26}$&--&1.39/21&0.73\\
  &        &diskbb&$0.20^{+0.20}_{-0.16}$&--&$2.73^{+1.14}_{-0.62}$&1.16/21&0.53\\
  &2004 August&pow&$0.55^{+0.15}_{-0.13}$&$1.69^{+0.20}_{-0.18}$&--&1.35/42&0.73\\
  &        &diskbb&$0.30^{+0.08}_{-0.07}$&--&$1.76^{+0.26}_{-0.21}$&1.18/42&0.87\\
  \hline
19$^{\xi}$&2001 February&pow&$0.57^{+0.06}_{-0.05}$&$1.67^{+0.09}_{-0.08}$&--&0.77/140&8.27\\
	&        &diskbb&$0.31^{+0.04}_{-0.03}$&--&$1.97^{+0.15}_{-0.13}$&1.04/140&7.50\\
  &2004 February&pow&$0.83^{+0.03}_{-0.02}$&$2.00^{+0.03}_{-0.03}$&--&1.05/628&18.40\\
  &        &diskbb&$0.48^{+0.01}_{-0.01}$&--&$1.57^{+0.03}_{-0.03}$&1.28/628&16.70\\
  &2004 August&pow&$0.62^{+0.03}_{-0.03}$&$1.81^{+0.05}_{-0.04}$&--&1.17/390&9.10\\
  &        &diskbb&$0.35^{+0.02}_{-0.02}$&--&$1.75^{+0.06}_{-0.06}$&1.76/390&8.20\\
  &2005 February&pow&$0.81^{+0.04}_{-0.04}$&$1.86^{+0.05}_{-0.05}$&--&0.91/331&23.30\\
  &        &diskbb&$0.47^{+0.02}_{-0.02}$&--&$1.77^{+0.07}_{-0.06}$&1.09/331&21.30\\
  &     &RS &$0.67^{+0.03}_{-0.03}$&--&$7.28^{+0.78}_{-0.62}$&0.88/330&22.60\\
  \hline
25$^{\xi}$&2001 February&pow&$2.36^{+0.35}_{-0.31}$&$1.84^{+0.19}_{-0.18}$&--&1.14/73&7.10\\
  &     &diskbb&$1.61^{+0.02}_{-0.20}$&--&$2.11^{+0.27}_{-0.22}$&1.05/73&6.50\\
  &2004 February&pow&$2.45^{+0.10}_{-0.10}$&$1.68^{+0.05}_{-0.05}$&--&1.36/670&33.30\\
  &        &disbb&$1.72^{+0.06}_{-0.06}$&--&$2.55^{+0.09}_{-0.08}$&1.18/670&31.70\\
  &2004 August&pow&$1.66^{+0.12}_{-0.12}$&$1.30^{+0.07}_{-0.07}$&--&1.08/310&8.90\\
  &        &diskbb&$1.28^{+0.08}_{-0.09}$&--&$3.40^{+0.27}_{-0.23}$&0.96/310&8.50\\
  &2005 February&pow&$1.87^{+0.32}_{-0.29}$&$1.56^{+0.17}_{-0.16}$&--&0.97/80&8.40\\
  &    &diskbb&$1.27^{+0.17}_{-0.19}$&--&$2.71^{+0.40}_{-0.31}$&0.94/80&8.00\\
  \hline
36&2004 February&pow+ga&$0.66^{+0.46}_{-0.06}$&$6.79^{-6.83}_{-0.08}$&$0.90^{+0.08}_{-0.08}$&0.86/30&0.15\\
  &             &bb+ga &$0.11^{+0.20}_{-0.10}$&--&$0.19^{+0.06}_{-0.04},0.94^{+0.04}_{-0.04}$&0.94/30&0.16\\
  &2004 August&pow+ga&$0.43^{+0.05}_{-0.04}$&$3.64^{+0.22}_{-0.23}$&$0.88^{+0.04}_{-0.07}$&0.97/116&0.86\\
  &             &bb+ga &$0.45^{+0.10}_{-0.06}$&--&$0.15^{+0.01}_{-0.02},0.0^{+0.87}_{-0.0}$&1.23/116&0.90\\
  \hline
44&2001 February&diskbb&$0.23^{+0.12}_{-0.11}$&--&$1.49^{+0.30}_{-0.22}$&0.49/32&1.40\\
  &2004 February&diskbb&$0.21^{+0.10}_{-0.09}$&--&$1.30^{+0.22}_{-0.17}$&0.73/31&0.63\\
  &2004 August&diskbb&$0.21^{+0.05}_{-0.05}$&--&$1.44^{+0.10}_{-0.14}$&0.97/60&0.85\\
  &2005 February&diskbb&$0.21^{+0.18}_{-0.15}$&--&$2.13^{+1.04}_{-0.52}$&1.00/16&1.50\\
  \hline
52&2001 February&pow&$0.54^{+0.43}_{-0.34}$&$2.84^{+1.00}_{-0.79}$&--&1.06/15&0.21\\
  &2004 February&pow&$0.55^{+0.18}_{-0.14}$&$3.36^{+0.63}_{-0.51}$&--&1.24/33&0.73\\
  &2004 August&pow&$0.73^{+0.20}_{-0.18}$&$3.52^{+0.48}_{-0.42}$&--&1.18/43&0.30\\
  &        &RS&$0.59^{+0.34}_{-0.23}$&--&$1.05^{+0.36}_{-0.22}$&1.08/42&0.27\\
\cutinhead{Spectral variables$^{\star}$ }
23&2001 February&pow&$0.00^{+1.89}_{-0.00}$&$0.47^{+1.72}_{-0.73}$&--&0.21/8&0.30\\
  &2004 February&pow&$0.57^{+0.83}_{-0.54}$&$1.53^{+1.07}_{-0.46}$&--&0.68/14&0.12\\
  &2004 August&pow&$0.64^{+0.48}_{-0.35}$&$1.81^{+0.58}_{-0.49}$&--&0.79/35&0.20\\
  &2005 February&pow&$0.65^{+1.08}_{-0.05}$&$2.52^{+2.53}_{-1.42}$&--&0.78/15&0.13\\
  \hline
24&2001 February&pow&$0.88^{+0.22}_{-0.40}$&$8.46^{-8.46}_{-2.80}$&--&0.94/38&0.09\\
  &2004 February&pow&$0.23^{+0.13}_{-0.10}$&$3.94^{+0.94}_{-0.70}$&--&1.52/36&0.12\\
24 (con't)  &        &RS&$0.010^{+0.03}_{-0.03}$&--&$0.83^{+0.07}_{-0.07}$&1.28/37&0.08\\
  &2004 August&pow&$0.55^{+0.18}_{-0.14}$&$6.34^{+1.26}_{-0.99}$&--&1.89/53&0.09\\
  &        &BB&$0.11^{+0.06}_{-0.03}$&--&$0.17^{+0.02}_{-0.02}$&1.67/53&0.09\\
  &2005 February&pow&$0.32^{+0.16}_{-0.12}$&$4.10^{+1.09}_{-0.83}$&--&0.95/32&0.21\\
  \hline
38&2001 February&pow+BB+ga&$0.57^{+0.33}_{-0.17}$&$3.07^{+0.05}_{-0.70}$&$0.95^{+0.95}_{-0.95},0.64^{+0.23}_{-0.42}$&1.12/99&1.56\\
  &        &pow+mekal&$0.72^{+0.17}_{-0.21}$&$2.63^{+0.18}_{-0.16}$&$0.24^{+0.11}_{-0.05}$&1.09/101&1.61\\
  &2004 February&pow+BB+ga&$0.55^{+0.60}_{-0.18}$&$2.58^{+0.49}_{-0.63}$&$0.17^{+0.06}_{-0.05},0.72^{+0.07}_{-0.45}$&1.41/88&1.32\\
  &        &pow+mekal&$0.43^{+0.13}_{-0.29}$&$2.19^{+0.25}_{-0.25}$&$0.49^{+0.13}_{-0.05}$&1.15/90&0.91\\
  &2004 August&pow+BB+ga&$0.30^{+0.11}_{-0.08}$&$1.66^{+0.34}_{-0.45}$&$0.23^{+0.04}_{-0.04},0.82^{+0.03}_{-0.05}$&1.26/113&1.15\\
  &        &pow+mekal&$0.35^{+0.04}_{-0.04}$&$1.30^{+0.21}_{-0.21}$&$0.61^{+0.04}_{-0.04}$&1.18/115&1.11\\
  &2005 February&pow+BB+ga&$0.27^{+0.15}_{-0.15}$&$1.00^{+1.38}_{-2.05}$&$0.28^{+0.10}_{-0.07},0.79^{+0.49}_{-0.36}$&1.26/44&1.57\\
  &        &pow+mekal&$0.38^{+0.60}_{-0.08}$&$1.50^{+1.16}_{-1.15}$&$0.55^{+0.08}_{-0.30}$&1.28/46&1.15\\
\cutinhead{Other bright X-ray sources}
21&2001 February&pow&$0.70^{+0.68}_{-0.53}$&$1.77^{+0.63}_{-0.54}$&--&0.81/11&0.48\\
  &2004 February&pow&$0.83^{+0.49}_{-0.23}$&$2.00^{+0.42}_{-0.33}$&--&0.60/14&0.35\\
  &2004 August&pow&$0.65^{+0.24}_{-0.20}$&$1.68^{+0.23}_{-0.21}$&--&0.87/26&0.54\\
  &        &diskbb&$0.34^{+0.15}_{-0.13}$&--&$2.01^{+0.38}_{-0.29}$&0.62/26&0.50\\
  &2005 February&pow&$0.79^{+0.84}_{-0.60}$&$2.27^{+1.13}_{-0.87}$&--&0.49/12&0.45\\
  &     &diskbb&$0.39^{+0.54}_{-0.20}$&--&$1.17^{+0.96}_{-0.43}$&0.47/12&0.37\\
  \hline
27&2004 February&NEI&$1.04^{+0.14}_{-0.14}$&$9.56^{+0.01}_{-0.01}$&$2.64^{+1.12}_{-1.12}$&1.29/14&0.07\\
  &2004 August&NEI&$0.81^{+0.13}_{-0.13}$&$9.11^{+0.01}_{-0.01}$&$2.11^{+0.56}_{-0.56}$&0.30/15&0.10\\
\enddata
\tablecomments{This table lists spectral parameters to energy spectra of sources with enough counts (over 100) in a particular observation for fitting. The sources in a particular group (i.e. flux, spectral variables, and other bright sources) are listed in ascending order of RA. For the XSPEC model, pow refers to power-law; BB refers to blackbody; diskbb refers to disk blackbody; RS refers to Raymond-Smith; ga refers to Gaussian lines. The quoted errors are at 90\% confidence level as generated by XSPEC.\\
$^{a}$This column shows the photon index ($\Gamma$) and the ionization timescale ($n_{e}t$) for power-law and NEI models respectively.\\
$^{b}$The temperature refers to blackbody temperature for blackbody, temperature at inner disk radius for disk blackbody, plasma temperature for Raymond-Smith and NEI, and line energy in keV for Gaussian. In the case of source 36 and 38, the 2 temperatures are blackbody temperature and gaussian line energy in this order.\\
$^{c}$ The quoted luminosities are absorbed value. The assumed distance to IC342 is 1.8 Mpc.\\
$^{\dagger}$ Sources with significant flux variability (i.e. $S_{\rm flux}\ge3$)\\
$^{\star}$ Sources with significant spectral variability that exhibit changes in the best-fitted spectral parameters at 90\% confidence level.\\
$^{\xi}$ These two sources are also classified as spectral variables.
\label{t:spe}}
\end{deluxetable}

\end{document}